\documentclass[axodraw,prd,showpacs,epsf,graphicx,preprintnumbers]{revtex4}

\usepackage{amssymb}
\usepackage{axodraw}
\usepackage{graphicx}
\usepackage{dcolumn}
\usepackage{bm}
\usepackage[cp1251]{inputenc}
\usepackage[T2A]{fontenc}
\usepackage[russian]{babel}
\usepackage{amsmath}
\usepackage{feyn}
\usepackage{feynmf}

\newcommand{\btr}{\bigtriangledown}
\newcommand{\sial}{\sin\alpha}

\newcommand{\coal}{\cos\alpha}

\newcommand{\la}{\lambda}
\newcommand{\lb}{\label}
\newcommand{\om}{\omega}
\newcommand{\al}{\alpha}
\newcommand{\bt}{\beta}
\newcommand{\ka}{\kappa}
\newcommand{\pa}{\partial}
\newcommand{\ga}{\gamma}
\newcommand{\si}{\sigma}
\newcommand{\Si}{\Sigma}

\newcommand{\fr}{\frac}
\newcommand{\de}{\delta}
\newcommand{\ff}{{f}}
\newcommand{\bw}{\begin{widetext}}
\newcommand{\ew}{\end{widetext}}
\newcommand{\be}{\begin{equation}}
\newcommand{\ee}{\end{equation}}
\newcommand{\ba}{\begin{eqnarray}}
\newcommand{\ea}{\end{eqnarray}}
\newcommand{\non}{\nonumber}
\newcommand{\va}{\varepsilon}
\newcommand{\ep}{\epsilon}

\newcommand{\st}{\stackrel}

\def\cF{{\cal F}}

\def\z{{\bf z}}

\def\e{{\rm e}}

\begin{document}
\title{Cerenkov radiation from moving  straight strings}
\author{D.V. Gal'tsov$^*$,  E.Yu. Melkumova$^*$ and K. Salehi$^*$}
\address{$^*$Department of Physics, Moscow State University,119899,
Moscow, Russia}


\begin{abstract}
We study Cerenkov radiation from moving straight strings which
glisse with respect to each other in such a way that the projected
intersection point moves faster than light. To calculate this
effect we develop  classical perturbation theory for the system of
Nambu-Goto strings interacting with dilaton, two-form and gravity.
In the first order  one encounters divergent self-action terms
which are eliminated by  classical renormalization of the string
tension. Cerenkov radiation arises in the second  order. It is
generated by an effective source which contains contributions
localized on the strings world-sheets and bulk contributions
quadratic in the first order fields. In the ultra-relativistic
limit  radiation exhibits angular peaking on the Cerenkov  cone in
the forward direction  of the fast string in the rest frame of
another. The radiation spectrum then extends up to high
frequencies proportional to square of the Lorentz-factor of the
relative velocity. Gravitational radiation is absent since the 1+2
space-time transverse to the straight string does not allow for
gravitons. A rough estimate of the Cerenkov radiation in the
cosmological cosmic strings network is presented.

\end{abstract}
\maketitle
\section{Introduction}
During past few years the hypothesis of cosmic strings received
new impetus from superstring theory. Although perturbative
superstrings in ten dimensions are too heavy to be admitted   as
cosmic strings, it was realized that there are various
possibilities for geometry of  string compactifications to
accomodate four-dimensional strings  with much lower tensions.
Copious production of cosmic strings is typical \cite{JO02,ST02}
for the brane inflation scenario \cite{DT99}, in which the period
of inflation is associated with the collision of branes.  This
scenario  provides an acceptable  model of inflation and predicts
creation of cosmic superstrings  consistent with the current CMB
data.  These strings typically have lower tensions than the usual
GUT cosmic strings and thus they are not the main players in the
formation of cosmic structures, but their observational signatures
could provide a direct confirmation of the string theory. This
stimulated a detailed study of creation and evolution of the
cosmic superstring network within the  KKLMMT model \cite{Ka03},
racetrack inflation \cite{Bl04} and other particular scenarios
\cite{Al02,FiTy05,ShTy06} (for recent review see
\cite{Po04,DaKi05,Cl05,Vi05,Sa06}). One particular feature of the
warped IIB compactifications involved in these considerations is
prediction of two types of cosmic strings: F-strings (fundamental)
and D-strings (Dirichlet) with different tensions $\mu_F,\,\mu_D$
\cite{DV04,CMP04,P04,JaJoPo04}. F and D strings may also form the
so-called $p,q-$ composites \cite{Sa05} and provide for triple
junctions \cite{CoKiSt06} with new exotic observational
predictions. Typical values of the dimensionless parameter $G\mu$
are in the interval $(10^{-11}-10^{-7})$. Another new feature is
that cosmic superstrings generically have lower reconnection
probabilities lying in the range $P\sim (10^{-3} - 1)$. This
changes cosmological evolution of the string network
\cite{Sa04,Ma04,CoSa05,AS05,Ty05} leading in particular to
enhancement of the fraction of straight strings.

Both the string network evolution and their possible observational
signatures crucially depend on radiation processes. In fact, it
was recognized long ago that oscillating loops of cosmic strings
generate large output of gravitational waves
\cite{VaVi85,Bu85,Vi85,GaVa87} at the level accessible for current
and future detectors. Global strings produce massless axions
\cite{D85,ViVa85a,Da85,GaVa87,DaQu89,CoHaHi90} which become
massive at a later stage of expansion, creating an observational
constraint on the axion mass. In the models containing  the
massless dilaton (like cosmic superstrings), the dilaton radiation
from strings may also constrain the string tension parameter
$G\mu$ \cite{GaVa87gb,DaVi97,BaKa05}.

The main mechanisms of radiation in the string network which have
been explored so far were  radiation from smoothly oscillating
string loops and radiation from kinks and cusps formed on them. It
was tacitly assumed that straight unexcited long strings do not
radiate. Meanwhile, interaction of  long strings via massless
fields gives rise to another radiation mechanism of Cerenkov
nature. When two straight non-excited  Nambu-Goto strings
interacting at a distance via dilaton, two-form and gravity glisse
with respect to each other, they get deformed in the vicinity of
the point of their minimal separation. This point can propagate
with  a faster-than light velocity, provided the inclination angle
between the strings is sufficiently small. (For the strictly
parallel strings this velocity is infinite for any relative string
speed.) In this case, the propagating deformation together with
the associated field tensions become the source of Cerenkov
radiation, which is the effect of the second order in interaction
of the string with massless fields. This mechanism was suggested
in \cite{GaGrLe93} for gravitationally interacting strings, but it
turned out that, although the effect is kinematically allowed, the
corresponding amplitude is zero on the mass-shell of the graviton.
The reason is that the $1+2$ space-time orthogonal to the straight
string  does note allow for gravitons. (However, gravitons {\em
are} produced at quantum level in the strings recombination and
annihilation \cite{CoPaSt06}.) But Cerenkov mechanism works for
other possible string massless excitations leading, e.g., to
electromagnetic radiation from superconducting strings
\cite{GaGrLa94} or emission of axions. The latter efect was
recently studied in detail in the flat space-time \cite{GMK04}. It
was found that, although being of the second order in the axion
coupling constant, it still gives a considerable contribution into
the total cosmological axion production by the string network.

Cerenkov radiation from straight strings is similar to
bremsstrahlung of point charges in electrodynamics. Moreover,  the
system of parallel strings interacting via a two-form field in D
space-time dimensions is exactly equivalent to the system of point
charges interacting with the vector field in (D-1)-dimensional
space-time.  A distinction  of the Cerenkov mechanism of
dissipation in the cosmic string network from conventional
radiation via formation of loops can be understood as follows.
Formation of loops from initially disconnected straight strings is
effected via  direct {\em contact} interaction of intersecting
strings. In our case the {\em long-range} interaction of strings
via massless fields which can be potentially radiated   underlies
the formation of the superluminal radiation source. Cerenkov
radiation is a higher order effect and thus potentially smaller.
But  it works for a wider set of initial data in the string
network (non-zero impact parameters of colliding strings), than
that corresponding to intercommuting strings. Also,as we will show
here, Cerenkov radiation is strongly enhanced in the case of
relativistic velocities. So it can still be non-small in the
cosmological setting. The detailed cosmological applications
remain outside the scope of the present paper, but we give the
rough estimates of the dilaton and two-form field backgrounds
generated via Cerenkov mechanism in the evolving string network.

In this paper we consider  Cerenkov radiation of moving straight
strings interacting with three massless fields: dilaton,
antisymmetric second rank tensor (NS-NS or RR two-forms) and
gravity. To facilitate construction of the solution of the coupled
system of the Nambu-Goto and the field equations in the second
order  we  introduce a diagrammatic representation similar to
Feynmann graphs. In the first order approximation one encounters
divergencies  which are cured  by classical renormalization of the
string tension. Then the effective sources of the dilaton and
two-form radiation are constructed  and the radiation rates are
calculated.

The plan of the paper is as follows. In Sec.~2 we give the general
setting of the problem and introduce graphic representation for
classical vertices corresponding to interactions of  strings with
dilaton, two-form and gravity  and to interactions between the
fields. Sec.~3 contains formulation of the perturbation approach
and an explicit construction of the second order equations. In
Sec.~4 we reproduce in our framework the main features of
classical renormalization for strings  (in the lowest order of
perturbation theory): elimination of  scalar and two-form
divergences via renormalization of the string tension, and the
absence of gravitational divergency. In Sec.~5 the first order
deformations of the string world-sheets due to interaction via the
dilaton, two-form and linearized gravity are considered. The main
calculation is presented in  Sec.~6 where we construct radiation
generating currents by computing the source terms in the wave
equations of the second order. These include the world-sheet local
terms and the bulk terms which correspond to multiple second order
graphs including all relevant elementary vertices. Radiation rates
for the dilaton and the two-form fields are computed in the Sec.~7
and analyzed in detail in the case of ultrarelativistic
velocities. Finally, in Sec.~8 we present rough cosmological
estimates and summarize our results in Sec.~9.

\section{Action and equations of motion }
We consider a pair $n=1,2$ of straight Nambu-Goto strings
described by the world-sheets\begin{equation}\label{ }
    x^{\mu}=X_n^{\mu}(\sigma^a_n),\;\; \mu=0,1,2,3,\;\; \sigma_a=(\tau,
\sigma),\;\; a=0,1.
\end{equation}
Strings interact via the gravitational $ g_{\mu\nu} \equiv
\eta_{\mu\nu} + h_{\mu\nu} $, the dilaton $ \phi (x) $ and the
two-form (axion)  $B_{\mu\nu}(x)$. Using the Polyakov form for the
string action, we present the total action in the form:
\ba\label{mac} S&=&-\sum\int\left\{
 \fr{\mu}{2} X^\mu_a X^\nu_b g
 _{\mu\nu}\gamma^{ab}\sqrt{-\gamma}\e^{2\al\phi}+2\pi\ff
 X^\mu_aX^\nu_b \epsilon^{ab} B_{\mu\nu}\right\}d^2 \sigma +
\non\\
&+&\int\limits\left\{ 2\partial_\mu\phi\partial_\nu\phi
g_{\mu\nu}+\fr{1}{6}H_{ \mu \nu \rho} H^{ \mu \nu
\rho}\e^{-4\al\phi} - \frac{R}{16\pi G } \right\}\sqrt{-g} \,
d^4x, \ea where the sum is taken over $n=1,2$ (the corresponding
index is omitted for brevity).   Our signature choice is mostly
minus for the space-time metric, and $(+,-)$ for the world-sheets.
The Levi-Civit\`a symbol is $\epsilon^{01}=-\epsilon^{10}=1,\;
\ga_{ab}$ is the metric on the world-sheet, and the lower Latin
indices mean partial derivatives with respect to the world-sheet
coordinates $X^\mu_a\equiv\partial_a X^\mu $. The action contains
four parameters: the string tension $\mu $, the Newton constant
$G$, the dilaton  coupling $\al$ and the two-form coupling $\ff$.
The antisymmetric three-form field strength is defined as \be H_{
\mu \nu \la}=\partial_\mu B_{\nu\la}+\partial_\nu
B_{\la\mu}+\partial_\la B_{\mu\nu}.\ee
\par
Variation of the action (\ref{mac}) with respect to $ X^\mu$ leads
to the equations of motion for strings  \be\lb{em}
\partial_a\left(\mu X_b^\nu
g_{\mu\nu}\ga^{ab}\sqrt{-\ga}\e^{2\al\phi}+4\pi f
X_b^\nu\epsilon^{ab}B_{\mu\nu}\right)=X^\la_a
X^\nu_b\partial_\mu\left(\fr{\mu}{2}
g_{\la\nu}\gamma^{ab}\sqrt{-\gamma}\e^{2\al\phi}+2\pi\ff
\epsilon^{ab} B_{\la\nu}\right).\ee The coordinate derivatives on
the right hand side apply to the metric, dilaton and two-form
fields, their values are taken on the world sheet. The derivatives
with respect to the world-sheet coordinates $\sigma^a$ on the left
hand side apply both to the world-sheet quantities
$X^\nu(\sigma^a),\,\gamma_{ab}$ and to  the metric, dilaton and
two-form fields, e.g., for the metric
\begin{equation}\label{ } \partial_a g_{\mu\nu}=
X^\lambda_a\partial_\lambda g_{\mu\nu},
\end{equation} and similarly for $\phi,\;B_{\mu\nu}$. In this
formalism $\gamma_{ab}$ is an independent variable; variation of
the action with respect to $\gamma_{ab}$ gives the constraint
equation\be\label{con} (X_a^\mu X_b^\nu - \fr12 \ga_{ab}\ga^{cd}
X_c^\mu X_d^\nu )g_{\mu\nu}=0,\ee whose solution defines
$\ga_{ab}$ as the induced metric on the world-sheet:
\be\label{ceq} \ga_{ab}=X_a^\mu X_b^\nu g_{\mu\nu}. \ee
\par Consider now the field equations. Variation over  $\phi$
gives the dilaton equation
\be\label{fe}\partial_\mu\left(g^{\mu\nu}\partial_\nu
\phi\sqrt{-g}\right)+\fr{\al}6
H^2\e^{-4\al\phi}=-\sum\fr{\mu\al}{4}\int X^\mu_a X^\nu_b
g_{\mu\nu}\ga^{ab}\sqrt{-\ga}\e^{2\al\phi}
\delta^4(x-X(\sigma,\tau))d^2\sigma,\ee where the sum in the
source term is taken over the contribution of two stings.  The
equation for the two-form field reads: \be\lb{fea}
\partial_\mu\left(H^{\mu\nu\la}\e^{-4\al\phi}\sqrt{-g}\right)=-\sum 2\pi
f \int X^\nu_aX^\la_b
\epsilon^{ab}\delta^4(x-X(\sigma,\tau))d^2\sigma.\ee We also have
the Bianchi identity \begin{equation}\label{ } \nabla_{[\mu}
H_{\al\beta\ga ]}=0,
\end{equation}
where  alternation over indices has to be performed and the
derivative can be equivalently treated as a covariant derivation
with respect to $g_{\mu\nu}$ or a partial derivative.
\par Finally, for the metric we have the Einstein equations
\be\label{gre} R_{\mu\nu}-\fr12g_{\mu\nu}R=8\pi
G(\stackrel{\phi}{T}{\!\!}_{\mu\nu}+
\stackrel{B}{T}{\!\!}_{\mu\nu}+\stackrel{st}{T}{\!\!}_{\mu\nu}),\ee
where the source terms read \ba\label{ts}
\stackrel{st}{T}{\!\!}^{\mu\nu}&=&\sum \mu\int X^\mu_a X^\nu_b
\gamma^{ab}\sqrt{-\gamma}\e^{2\al\phi}
\fr{\delta^4(x-X(\sigma,\tau))}{\sqrt{-g}}d^2\sigma,\\ \label{tp}
\stackrel{\phi}{T}{\!\!}^{\mu\nu}&=&4\left(\pa^\mu\phi\partial^\nu\phi-\fr12
g^{\mu\nu}(\btr\phi)^2\right), \\ \label{tb}
\stackrel{B}{T}{\!\!}^{\mu\nu}&=&\left(H^\mu_{\al\bt}H^{\nu\al\bt}-
\fr16H^2g^{\mu\nu}\right)\e^{-4\al\phi}.\ea
The total system of equations consists of two equations for stings
of the type (\ref{em}) and the field equations (\ref{fe},
\ref{fea}, \ref{gre}). It  describes classically in a
self-consistent way   motion of the strings with account for their
interaction via the dilaton, two-form and gravitational fields as
well as   evolution of the generated fields. This system can be
solved iteratively in terms of the coupling constants $\al, \ff,
G$. To facilitate this construction it is convenient to introduce
a classical analog of the Feynmann graphs, denoting string by bald
solid line, the dilaton by this solid line, the two-form by dashed
and graviton by wavy lines (Fig. 1). Simple analysis of the
equations reveals that we have vertices involving linear and
non-linear string-field interactions, as well as three-leg and
multi-leg interactions of fields between themselves.
\begin{figure}
\centering
\includegraphics[width=6cm,height=7cm]{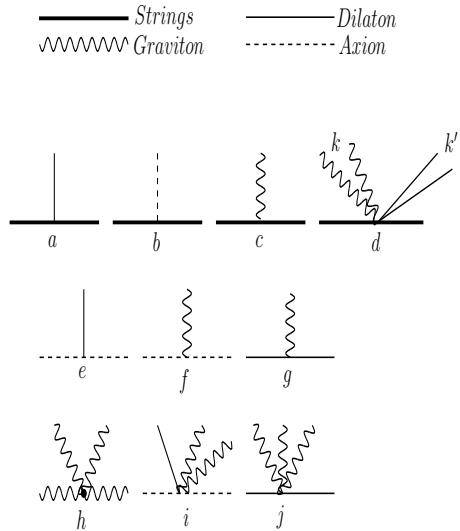}
\caption{Vertices associated with the action (\ref{mac}): graphs
$a,b,c$ are string-dilaton, string-two-form and string-graviton
vertices of the lowest order; graph $d$ depicts higher-order
string-dilaton-graviton vertices (for all $k\geq 1, k'\geq 1$);
graphs $e,f,g$ are lowest order field interactions present in the
action (\ref{mac}), and graphs $e,f,g$ are multi-graviton vertices
accompanying the lowest order ones.} \label{f:1}
\end{figure}
\section{Perturbation theory}
We construct an iterative solution of the coupled string-field
system expanding  the string world-sheet mapping functions
$X^\mu_n(\sigma^a),\; n=1,2,$ and the field variables in terms of
the coupling constants $\al, \ff, G$:
 \ba \label{expa} X^\mu &=&\st{0}X{\!\!}^\mu
+\st{1}X{\!\!}^\mu+\ldots,\non\\
\phi&=& \stackrel{1}{\phi} + \stackrel{2}{\phi}+ \ldots,
\non\\B_{\mu\nu} &=& \stackrel{1}{B}{\!\!}_{\mu\nu} +
\stackrel{2}{B}{\!\!}_{\mu\nu}+ \ldots ,\non\\ h_{\mu\nu} &=&
\stackrel{1}{h}{\!\!}_{\mu\nu} + \stackrel{2}{h}{\!\!}_{\mu\nu}+
\ldots . \ea Here the expansions of $X^\mu_n $ start from the zero
order, while those of the field variables  --- from the first
order terms, that is we assume that there are no background
dilaton, two-form or gravitational fields. Zero order mapping
functions describe the straight infinite uniformly moving strings
 \be\label{xo}
   \stackrel{0}{X}{\!\!}^\mu_n \ = \ d^\mu_n \ + \ u^\mu_n \tau \
+ \Sigma^\mu_n \sigma.   \ee Here $ \Sigma^\mu_n $ is the unit
spacelike constant vectors along the strings, and $ u^{\mu}_n$ are
the unit timelike   constant  vectors --- four-velocities of the
strings. The corresponding three-dimensional velocities are
orthogonal to the strings. The constant vectors $ d^\mu_n $ can be
regarded as impact parameters for two strings with respect to the
chosen frame. In the zero order the space-time metric is flat, and
the corresponding world-sheet metrics are also Minkowskian
$\eta_{ab}={\rm diag}(1,-1)$ in view of the normalization assumed
\be\label{orto}
(\Sigma_n\Sigma_n)=\eta_{\mu\nu}\Sigma^\mu_n\Sigma^\nu_n=-1, \quad
(u_n u_n) =1,\quad  (\Sigma_n u_n)=0.\ee It is convenient the
choose the Lorentz frame in which the first string is at rest and
is stretched along the z-axis: \be\lb{usig}
u^\mu_1=[1,0,0,0],\quad \Sigma^\mu_1=[0,0,0,1].\ee The second
string is assumed to move in the plane $x^2,\,x^3$ with the
velocity  $ v $ orthogonal to the string itself: \be u^\mu_2=\ga
\, [1,0,-v\coal,v\sial],\quad \Sigma^\mu_2=[0,0,\sial,\coal],\ee
where $\ga  = (1-v^2)^{- 1/2} $. In such a configuration the
strings never intersect each other remaining always in the
parallel planes. Apart from the orthogonality conditions
(\ref{orto}), four other scalar products are
\begin{equation}\label{scprod}
(u_1,u_2)=\gamma,\quad
(\Sigma_1,\Sigma_2)=-\coal,\quad(u_1,\Sigma_2)=0,\quad
(u_2,\Sigma_1)=-v\gamma\sial,\quad
\end{equation} note that $u_1$ and $\Sigma_2$ are orthogonal.
We also choose both impact parameters $d^\mu_n$ to be orthogonal
to $u^\mu_n$ and $\Sigma^\mu_n$ and aligned with the axis $x^1$,
the distance between the planes being $d=d_2-d_1$. The angle of
inclination $\al$ of the second string with respect to the first
one can be written in a Lorentz-invariant form  \be\label{ang}
\alpha = \arccos(- \Sigma_1 \Sigma_2). \ee Similarly, the
invariant expression for the relative velocity of the strings is
\be \lb{vel} v \ = \ (1-(u_1u_2)^{-2})^{ \frac {1}{2}}. \ee With
this parametrization of the unperturbed world-sheets, the
projected intersection point (the point of the minimal separation
between the strings) moves with the velocity \be\lb{vp} v_p=
\frac{v}{\sin \alpha}=(u_1 u_2)^{-1}\left( \frac{(u_1u_2)^2-1}
  {1-( \Sigma_1 \Sigma_2)^2}\right)^{ \frac{1}{2}}
\ee along the $x^3$-axis. This motion is not associated with
propagation  of any signal, so the velocity $v_p$ may be
arbitrary, in particular, superluminal $v_p>1$. The case of
parallel strings corresponds to $v_p=\infty$.
\par
Note that the above introduced parameters (\ref{vel},\ref{ang})
are not invariant under reparameterizations of the world-sheets
\cite{GMK04}. The  quantity which is invariant under the volume
preserving reparameterizations is \be \kappa=\det\big( X_{1a}^\mu
X_{2b}^\nu \eta_{\mu\nu}\big)=\ga\coal.\ee The superluminal regime
corresponds to $\ka>1$.
\par
The expansions (\ref{expa}) are substituted into the system of
equations (\ref{em}, \ref{fe}, \ref{fea}, \ref{gre}) which has to
be solved  iteratively. The zero order differential operator in
the dilaton equation (\ref{em}) is the flat-space D'Alembert
operator $\Box=-\eta^{\mu\nu}\partial_\mu\partial_\nu$. Similarly,
choosing the Lorentz gauge for the two-from and the metric
perturbations \be\partial_\mu B^{\mu\nu}=0,\quad
\partial_\mu\psi^{\mu\nu}=0,\;\;\;
\psi^{\mu\nu}=h^{\mu\nu}-\frac12 \eta^{\mu\nu}h,\ee where
$h=h^\mu_\mu,$ we get the linear D'Alembert equations for the
first order two-form and the gravitational field as well. Due to
linearity of the field equations, the first order  dilaton,
two-form and metric perturbations can be presented as the sums of
the separate contributions due to two strings \ba\label{lph}
\stackrel{1}{\phi}&=&\stackrel{1}{\phi}{\!\!}_1
+\stackrel{1}{\phi}{\!\!}_2, \non\\
\stackrel{1}{B}{\!\!}^{\mu\nu}&=& \stackrel{1}{B}{\!\!}^{\mu\nu}_1
+ \stackrel{1}{B}{\!\!}^{\mu\nu}_2,  \\
\stackrel{1}{h}{\!\!}^{\mu\nu}&=&\stackrel{1}{h}{\!\!}^{\mu\nu}_1+
\stackrel{1}{h}{\!\!}^{\mu\nu}_2.\non \ea Here each term with
$n=1,2$ satisfies the individual D'Alembert equation with the
source labelled by the same index $n$: \ba
 \Box \, \stackrel{1}{\phi}_{n}&=&4\pi\,\stackrel{0}{J}_{n},
\lb{eeqop} \\
\Box \,\stackrel{1}{B}{\!\!}_{n}^{\mu\nu}&=&
4\pi\stackrel{0}{J}{\!\!}_{n}^{\mu\nu}, \lb{eeqoB}\\
\Box\,\stackrel{1}{h}{\!\!}^{\mu\nu}_n&=&4\pi
\stackrel{0}{\tau}{\!\!}^{\mu\nu}_n.\lb{feqh} \ea The coupling
constants are included into the source terms, while  zero indices
in the sources indicate that they  are computed using the zero
order approximations for the strings mapping functions. The source
terms thus read \ba \stackrel{0}{J}_{n }&=& \fr{\al\mu}{8\pi}
\int\limits \, \delta^4
\left(x-\stackrel{0}{X}_n(\tau,\sigma)\right)\,d^2\sigma, \lb{jop} \\
\lb{joB} \stackrel{0}{J}{\!\!}_{n }^{\mu\nu}&=& \fr{f}{2}
\int\limits\, {V}{\!\!}_{n}^{\mu \nu} \delta^4 \left(x-
\stackrel{0}{X}_(\tau,\sigma)\right)\,d^2\sigma ,\\ \lb{tau0}
\stackrel{0}{\tau}{\!\!}^{\mu\nu}_n&=& 4 G\mu_n\int W^{\mu\nu}_n
\delta^4 \left(x- \stackrel{0}{X}_n(\tau,\sigma)\right)
\,d^2\sigma, \ea where the following antisymmetric and symmetric
tensors are introduced \ba \lb{Vmn} V^{\mu\nu}_n&=&
\epsilon^{ab}\stackrel{0}{X}{\!\!}^{\mu}_{an}
\stackrel{0}{X}{\!\!}^{\nu}_{bn}=u^\mu_n\Sigma^\nu_n-
u^\nu_n\Sigma^\mu_n,\\ \label{u} U_n^{\mu \nu} &=& \eta^{ab}
\stackrel{0}{X}{\!\!}^{\mu}_{an} \stackrel{0}{X}{\!\!}^{\nu}_{bn}
=u^\mu_n u^\nu_n-\Sigma^\mu_n\Sigma^\nu_n , \quad
U_n^{\mu\nu}\eta_{\mu\nu}=U_n=2,\\\label{w}W^{\mu\nu}_n&=&U^{\mu\nu}_n-
\fr12\eta^{\mu\nu}U_n. \ea
\par Now consider the perturbations of the strings world-sheets induced
metrics (index $n$ is omitted) \be\lb{etamn}
\ga_{ab}=\eta_{ab}+\stackrel{1}{\ga}{\!\!}_{ab},\ee where in the
general gauge the first order correction reads
\be\lb{gab1}\stackrel{1}{\ga}{\!\!}_{ab}=
\stackrel{0}{X}{\!\!}^{\mu}_a \stackrel{0}{X}{\!\!}^{\nu}_b
\stackrel{1}{h}{\!\!}_{\mu\nu}
+2\stackrel{0}{X}{\!\!}^{\mu}_a\stackrel{1}{X}{\!\!}^{\nu}_b
\eta_{\mu\nu}.\ee However, to be able to disentangle the higher
order perturbed  equations one has to get rid of the second term
in (\ref{gab1}) proportional to $\stackrel{1}{X}{\!\!}^{\nu}$.
Using the space-time and the world-sheet diffeomorphism
invariance, we can impose the gauge condition \be\lb{cond}
\stackrel{0}{X}{\!\!}^{\mu}_a\stackrel{1}{X}{\!\!}^{\nu}_b
\eta_{\mu\nu}=0,\ee in which case the perturbed induced metric
will contain only the zero order string mapping functions:
\be\lb{sim}
\stackrel{1}{\ga}{\!\!}_{ab}=\stackrel{0}{X}{\!\!}^{\mu}_a
\stackrel{0}{X}{\!\!}^{\nu}_b \stackrel{1}h{\!\!}_{\mu\nu},\quad
\stackrel{1}{\ga}= U^{\mu\nu}\stackrel{1}h{\!\!}_{\mu\nu}.\ee
\par The first order perturbations of the mapping functions
$\stackrel{1}{X}{\!\!}^{\mu}$ describing deformations of the flat
world-sheets satisfy the  following equations following from the
Eq.(\ref{em}): \ba  &\mu\eta^{ab}\partial_a
\stackrel{1}{X}{\!\!}^{\nu}_b\eta_{\mu\nu}+\mu\partial_a
\left\{\stackrel{0}{X}{\!\!}^{\nu}_b
\left[\left(\stackrel{1}{h}{\!\!}_{\mu\nu}+\fr12\eta_{\mu\nu}
\stackrel{1}{\ga}\right)
\eta^{ab}-\eta_{\mu\nu}\stackrel{1}{\ga}{\!\!}^{ab}\right]+
2\al\eta_{\mu\nu}\eta^{ab}\phi\right\}+&\non\\
 & + 4\pi f \epsilon^{ab}\stackrel{0}{X}{\!\!}^{\nu}_b\,\partial_a
\!\stackrel{1}{B}{\!\!}^{\mu\nu}-2\al\mu
\partial_\mu\stackrel{1}{\phi}-\fr12 U^{\la\nu}
\partial_\mu\stackrel{1}{h}{\!\!}_{\la\nu}-2\pi f V^{\la\nu}\partial_\mu
\!\stackrel{1}{B}{\!\!}_{\la\nu}=0,\ea where raising of the
world-sheet indices is performed by the Minkowski metric:
\be\lb{gabup}
\stackrel{1}{\ga}{\!\!}^{ab}=\stackrel{1}{\ga}{\!\!}_{cd}\eta^{ac}\eta^{bd},\quad
\stackrel{1}{\ga}=\stackrel{1}{\ga}{\!\!}_{cd}\eta^{cd}.\ee
Differentiating  the field variables along the world-sheet
according to the rules \be
\partial_a\stackrel{1}{\phi}=\stackrel{0}{X}{\!\!}^{\la}\partial_\la
\stackrel{1}{\phi}, \quad \partial_a
\stackrel{1}{B}{\!\!}_{\mu\nu}
=\stackrel{0}{X}{\!\!}^{\la}_a\partial_\la
\stackrel{1}{B}{\!\!}_{\mu\nu}, \quad
\stackrel{1}{h}{\!\!}_{\mu\nu}
=\stackrel{0}{X}{\!\!}^{\la}_a\partial_\la
\stackrel{1}{h}{\!\!}_{\mu\nu},\ee  we can rewrite the above
equation as  \be\lb{eqmo}  \mu\eta^{ab}\partial_a\partial_b
\stackrel{1}{X}{\!\!}^{\mu}=F^\mu_{(\phi)}+F^\mu_{(B)}+F^\mu_{(h)},
\ee where the forces due to dilaton, two-form and graviton are
introduced: \ba\lb{fp}  F^\mu_{(\phi)}&=&\al\mu\left(U\partial^\mu
\stackrel{1}{\phi}-2U^{\mu\nu}\partial_\nu
\stackrel{1}{\phi}\right), \\ \lb{fb}  F^\mu_{(B)}&=&2\pi
fV^{\nu\la} \stackrel{1}{H}{\!\!}^\mu_{\;\nu\la}, \\ \lb{fh}
F^\mu_{(h)}&=&\mu\left[\stackrel{1}{h}{\!\!}_{\la\tau,\nu}
\left(U^{\mu\la} U^{\nu\tau}-\fr12U^{\mu\nu}U^{\la\tau}\right)-
U_{\nu\la}\left(\stackrel{1}{h}{\!\!}^{\mu\nu,\la}-\fr12\stackrel{1}
{h}{\!\!}^{\nu\la,\mu}\right)\right].\ea In these equations the
indices labelling the strings are not shown, but it is understood
that for each string we have to take into account at the right
hand side both the self-force terms arising form the fields due to
the same string, and the mutual interaction terms coming from the
partner string.
\par
Consider now the second order field equations. Expanding the Eqs.
(\ref{fe})-(\ref{gre}) to the next order and imposing again the
Lorentz gauge for the two-form and the linearized gravity  one
obtains the D'Alembert equations for the second order fields with
the source terms involving  contributions due to deformations of
the strings world-sheets (local terms) as well as the quadratic
combinations of the first order fields (bulk terms). The dilaton
equation with account for (\ref{sim}) will read \ba\lb{sod} \Box\;
\stackrel{2}{\phi}\;=\; 4\pi\stackrel{1}{J}&=& \fr{\mu\al}{4}\sum
\int \left[ {U}^{\mu\nu}\left(\stackrel{1}{h}_{\mu\nu}+
2\eta_{\mu\nu}\al\stackrel{1}{\phi}\right)-
4\stackrel{1}{X}{\!\!}^{\mu}\partial_\mu\right]\delta^4
\Big[x-\stackrel{0}{X}(\tau,\si)\Big]
 d^2\sigma + \non\\
&+&\fr\al6
\stackrel{1}{H}{\!\!}^{\mu\nu\la}\stackrel{1}{H}{\!\!}_{\mu\nu\la}
-\partial_\mu\left[\left(\stackrel{1}{h}{\!\!}^{\mu\nu}-
\fr12\eta^{\mu\nu}\stackrel{1}{h}\right)\partial_\nu
\stackrel{1}{\phi}\right],\ea where the partial derivative
operator acts on the delta function. Note that the first order
terms in the first line are multiplied by the dilaton coupling
constant and thus give the same order quantities as the products
of two first order field quantities.
\par
The equation for the second order two-form field is:
\ba\lb{soa}&\Box& \stackrel{2}{B}{\!\!}^{\mu\nu}=
4\pi\stackrel{1}{J}{\!\!}^{\mu\nu}\;=\non\\
&=& 2\pi f\sum \int \left(2\ep^{ab}\stackrel{0}{X}{\!\!}^{[\mu
}_a\stackrel{1}{X}{\!\!}^{\nu] }_b
\delta^4\Big[x-\stackrel{0}{X}(\tau,\si)\Big]-V^{\mu\nu}\stackrel{1}{X}{\!\!}^{\la
}\partial_\la\delta^4\Big[x-\stackrel{0}{X}(\tau,\si)\Big]\right)
d^2\sigma+\non\\&+&\partial_\la\left[\left(\fr12\stackrel{1}{h}-
4\al\stackrel{1}{\phi}\right)
\stackrel{1}{H}{\!\!}^{\la\mu\nu}\right],\ea where alternation
over indices includes the factor $1/2$.
\par The situation is slightly more complicated  for the graviton.
To obtain an equation for the second order gravitational
perturbation one has to take into account the quadratic terms in
the Einstein equations. In the gauge
$$\partial_\mu  {\psi}^{\mu\nu}=0$$ one has the following
expansion of the Einstein tensor up to the second order: \be
R_{\mu\nu}-\fr12g_{\mu\nu}R=\fr12 \Box {\psi}_{\mu\nu}+\fr12
S_{\mu\nu}(\stackrel{1}{h}{\!\!}_{\la\tau }),\ee where
$\psi_{\mu\nu}$ contains the first and the second order
quantities, and the quadratic term reads: \ba\lb{Smn}
S_{\mu\nu}&=&[\pa^\beta {h}_\mu^{\al}(\pa_\al h_{\nu\beta} -\pa_b
h_{\nu\al})-\fr12\pa_\mu h^{\al\beta}\pa_\nu
h_{\al\beta}-\fr12h_{\mu\nu}\pa_\al\pa^\al h + \non\\
&+&h^{\al\beta}(\pa_\nu\pa_\beta h_{\mu\al}+\pa_\mu\pa_\beta
h_{\nu\al}-\pa_\mu\pa_\nu h_{\al\beta}-\pa_\al\pa_\beta
h_{\mu\nu}) + \\& +&
\fr12\eta_{\mu\nu}(2h^{\al\beta}\pa^\la\pa_\la
h_{\al\beta}-\pa_\la h_{\al\beta}\pa_\beta h^{\al\la}+\fr32
\pa_\la h_{\al\beta}\pa^\la h^{\al\beta})].\non\ea Extracting the
second order terms we obtain from the Eq.(\ref{gre}): \be
\Box\stackrel{2}{\psi}{\!\!}_{\mu\nu}=16\pi G
\stackrel{1}\tau{\!\!}_{\mu\nu}, \ee where the source term reads:
 \ba \lb{taumn}\stackrel{1}\tau{\!\!}_{\mu\nu} &=&
\fr{\mu}2 \sum \int
\Big(2\stackrel{0}{X}{\!\!}^a_{(\mu}\stackrel{1}{X}{\!\!}_{\nu)\,a}
-\stackrel{1}{X}{\!\!}^{\lambda}
\partial_\lambda\Big)\delta^4(x-\stackrel{0}{X}(\tau,\si))d^2\si+
\non\\&+&\fr{\mu}4 \sum
\int\Big[\left(U_{\mu\nu}U^{\la\tau}-2U^\la_\mu U^\tau_\nu\right)
\stackrel{1}{h}_{\la\tau} + U_{\mu\nu}\Big(4\al\stackrel{1}{\phi}-
\stackrel{1}{h}\Big)\Big]\delta^4(x-\stackrel{0}{X}(\tau,\si))d^2\si+\\
&+&G^{-1}S_{\mu\nu}(\stackrel{1}{h}{\!\!}_{\la\tau})+2\pa_\mu
\stackrel{1}{\phi}\pa_\nu\stackrel{1}{\phi} -\eta_{\mu\nu}\pa_\la
\stackrel{1}{\phi}\pa^\la\stackrel{1}{\phi}+
\fr12\stackrel{1}{H}{\!\!}_{\mu\la\tau}\stackrel{1}{H}{\!\!}_\nu^{\;\la\tau}
-\fr1{12}\eta_{\mu\nu}\stackrel{1}{H}{\!\!}_{\nu\la\tau}
\stackrel{1}{H}{\!\!}^{\nu\la\tau}\non .\ea
\par All source terms in the second order field equations have similar
structure. Note  the presence of the derivatives from delta
functions in the strings world-sheet contributions. \par It is
worth noting that the second order dilaton, two-form and gravity
fields already can not be presented as a sum of terms due to
separate strings, but rather look as generated by the collective
sources. These sources contain contributions not only from the
perturbed world-sheets, but also the bulk contributions which are
not associated with separate strings. As we will see later, these
sources may have superluminal nature, in which case the Cerenkov
radiation will appear.

\section{Renormalization}
The action of the proper fields upon the source string is
described by the self-action terms in the Eq. (\ref{eqmo})
corresponding to the graphs shown on Fig. 2.
\begin{figure}
\centering
\includegraphics[width=6cm,height=1.2cm]{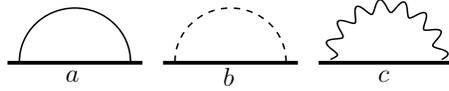}
\caption{Graphs describing self-interaction due to  dilaton ($a$),
two-form ($b$)  and linearized gravity ($c$). The contributions
from $a$ and $b$ are divergent and have different signs. The
contribution of $c$ is zero.}\label{f:2}
\end{figure}
the dilaton and the two-form lead to divergences while the
contribution from the linearized gravity  vanishes. Divergent
terms can be absorbed by renormalization of the string tension
parameter $\mu $ \cite{BaSh95,BaCar97}.  We consider
renormalization  in the first order of the perturbation theory.
Linearizing the string part of the action (\ref{mac}), one can
split it into the sum: \be\lb{S} S=S_{st}+S_{\phi}+S_B +S_h , \ee
where $ S_{st} $ is Polyakov action with the bare tension
parameter:
 \begin{equation}\label{Sst}
    S_{st}=-\fr{\mu_0}{ 2} \int X^\mu_a X^\nu_b
\eta_{\mu\nu}\eta^{ab} d^2\si,
\end{equation}
and three other terms describe interaction with the dilaton, the
two-form and the linearized gravity: \ba  \lb{Sphi}
S_{\phi}&=&-\mu_0 \al
\int\phi\; X^\mu_a X^\nu_b \eta_{\mu\nu}\eta^{ab}\; d^2\si,\\
\lb{SB} S_{B}&=&-2\pi f \int B_{\mu\nu}\; X^\mu_c X^\nu_d
\epsilon^{cd}
 \;d^2\si,\\  \lb{Sh} S_{h}&=&-\fr{\mu_0}{2} \int
X^\mu_a X^\nu_b \eta^{ab} h_{\mu\nu}\; d^2\si.\ea Since we are
working in the lowest order of the perturbation theory, the
mapping functions $X^\mu$ here are quantities of zero order. In
obtaining the last formula we  used the following expansion:
\be\lb{gabg} \ga^{ab}\sqrt{-\ga}=\eta^{ab}+h_{\mu\nu} \left(\fr12
X^\mu_c X^{\nu \, c} \eta^{ab}-X^{\mu \, a}X^{\nu \, b}\right)+
\cdots \ee Consider the first order dilaton field on the
world-sheet of the source string \be\lb{phst} \phi (\tau,\si)= \fr
{\al\mu_0}{8\pi^2}
\int\fr{\e^{iq_\mu(d^\mu-X^\mu(\tau,\si))}\delta(qu)\delta(q\Sigma)}{q^2}d^4
q,\ee where $qu=q_\mu u^\mu,\, q\Sigma=q_\mu\Sigma^\mu$ are the
flat-space scalar products, and $q^2=q_\mu q^\mu$. Due to
delta-functions, the scalar product in the exponential  is
constant: \be q_\mu
X^\mu(\tau,\si)=(qd)+(qu)\tau+(q\Sigma)\si=(qd)={\rm const}, \ee
so the integrand does not depend on $\si$ and $\tau$. The integral
diverges as \be \lb{Id} I=-\int\fr{\delta(qu)\delta(q\Sigma)}{q^2}
d^4 q \, = \,2\pi \int_0^\infty \fr{dq_{\perp}}{q_{\perp}}\ee
where we used the frame (\ref{usig}) and introduced polar
coordinates in the 2-plane orthogonal to $u^\mu$ and $\Sigma^\mu$:
\be\lb{q2} q^2=(qu)^2-(q\Sigma)^2-q_1^2-q_2^2, \quad
q_1^2+q_2^2=q_{\perp}^2. \ee The integral logarithmically diverges
at both ends. With the infrared (IR) an ultraviolet (UV) cut-off
parameters $q_{\perp}^{\rm min},\; q_{\perp}^{\rm max}$  one can
write:  \be \lb{Il} I=2\pi \ln{\fr{q_{\perp}^{\rm
max}}{q_{\perp}^{\rm min}}}.\ee Substituting this into the Eq.
(\ref{phst}) and further into (\ref{Sphi}) we find the regularized
dilaton part of the action: \be\lb{Sphir}
S_{\phi}^{reg}=\fr{\mu_0^2 \al^2 }{4\pi}\ln{\fr{q_{\perp}^{\rm
max}}{q_{\perp}^{\rm min}}}\int X^\mu_a X^\nu_b
\eta_{\mu\nu}\eta^{ab} d^2\si .\ee   Since the functional is the
same as the bare action functional $S_{st}$, one can absorb
divergencies by  renormalization of $\mu_0$.
\par
Similarly, the first order two-form field on the world-sheet of
the source string reads: \be\lb{Bst} B^{\mu\nu} (\tau,\si)= \fr
{f}{2\pi}
\int\fr{\e^{iq_\mu(d^\mu-X^\mu(\tau,\si))}V^{\mu\nu}\delta(qu)\delta(q\Sigma)}{q^2}d^4
q.\ee This integral also diverges as (\ref{Il}). In view of the
relation \be \lb{relV} V^{\mu\nu}V_\nu^\la=-U^{\mu\nu}, \ee one
can see that the action ({\ref{SB}) also have the functional form
of (\ref{Sst}), namely \be\lb{SBr} S_{B}^{reg}=-2\pi
f^2\ln{\fr{q_{\perp}^{\rm max}}{q_{\perp}^{\rm min}}}\int X^\mu_a
X^\nu_b \eta_{\mu\nu}\eta^{ab} d^2\si .\ee Finally for the
graviton $h_{\mu\nu}(\tau,\si)$ on the world-sheet we have the
divergent integral \be\lb{hst} h_{\mu\nu} (\tau,\si)= \fr{4 G
\mu}{\pi} \int \fr{\e^{iq_\mu(d^\mu-X^\mu(\tau,\si))}W_{\mu\nu}
\delta(qu)\delta(q\Sigma)}{q^2}\;d^4 q=-8G\mu W_{\mu\nu}
\ln{\fr{q_{\perp}^{\rm max}}{q_{\perp}^{\rm min}}},\ee where
$W_{\mu\nu}$ id given by the Eq. (\ref{w}). However, substituting
this into (\ref{Sh}) one obtains zero in view of the identity \be
W_{\mu\nu}U^{\mu\nu}=0. \ee Therefore, gravitational interaction
of the strings  does not lead to classical divergences in the
lowest order of the perturbation theory. This result is conformal
with previous results \cite{Car98,BuDa98} obtained with different
tools.
\par Collecting the above formulas, we   see that to remove
self-interaction divergences  one has merely to replace  the
tension parameter in the action (\ref{Sst}) as follows:
\be\lb{ten} \mu_0-\left( \fr{\mu_0^2\al^2}{2\pi}-4\pi f^2)\right)
\ln{\fr{q_{\perp}^{\rm max}}{q_{\perp}^{\rm min}}}=\mu.\ee
Divergences due to the dilaton and the two-form have opposite
signs. This reflects the fact that the scalar interaction is
attractive, while interaction via the two-form is repulsive. If
the Bogomolny-Prasad-Sommerfield (BPS) relation between the
dilaton and the two-form caouplings is satisfied \be\lb{BPS}
\fr{\mu^2\al^2}{2\pi}=4\pi f^2, \ee the divergent terms mutually
cancel, and there is no renormalization at all, $\mu=\mu_0$ (for
earlier work on this subject see refs.
\cite{BaSh95,BaCar97,Car98,BuDa98,Car98a,Gr01}. Note that our
dilaton coupling constant has dimension of length, the usual
dimensionless constant $\bar\alpha$ (quantity of the order of
unity) is related to it as \be\lb{albar}\alpha^2 =G
\bar\alpha^2.\ee \par It has to be noted that renormalization can
be performed in a simple way only at the linearized level. When
all non-linearities are taken into account, classical
renormalizability of the bosonic string theory interacting with
gravity is lost. In this paper we will be restricted by the second
order calculation of radiation which involves only on-shell second
order quantities. These are unaffected by the higher order
renormalization effects.  So, in what follows, we will use the
value for the string tension renormalized in the first-order, and
thus omit all self-interaction terms.
\section{Perturbations of the string world-sheets}
Now consider the first order perturbations of the mapping
functions $X^\mu_n(\tau,\si),\;n=1,2 $ caused by mutual
interactions. We have to substitute into each string equation of
motion (\ref{em}) the first order fields generated by another
string. It is convenient to use the following index convention:
$n=1,2,\; n'=1,2,\; n\neq n'$. The total perturbation thus splits
into three terms due to dilaton, two-from and graviton exchange:
\be\lb{X1}
\stackrel{1}{X}{\!\!}^\mu_n=\stackrel{\phi}{X}{\!\!}^\mu_n+
\stackrel{B}{X}{\!\!}^\mu_n+\stackrel{h}{X}{\!\!}^\mu_n, \ee  as
shown on Fig. 3. \begin{figure} \centering
\includegraphics[width=6cm,height=1.cm]{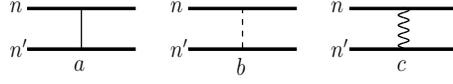}
\caption{Deformations of the strings world-sheets due to
interactions via the dilaton (a), two-form (b) and linearized
gravity (c); $n=1,2,\,n'=1,2,\,n\neq n'$.} \label{f:3}
\end{figure}Let us start with dilaton exchange contribution
(Fig. 3a). The corresponding world-sheet perturbation
$\stackrel{1}{X}{\!\!}^\mu_n (\tau,\si)$ is the solution of the
two-dimensional D'Alembert equation (following from the Eqs.
(\ref{eqmo},\ref{fp})): \be\lb{em11} (\pa^2_\tau
-\pa^2_\si)\stackrel{\phi}{X}{\!\!}^\mu_n =
-2\al\left[(\stackrel{\phi}{U}{\!\!}^{\mu\nu}\pa_\nu-\pa^\mu)
\stackrel{1}{\phi}(x)\right]_{x=\stackrel{0}X_n(\tau,\si )}\, ,\ee
where the dilaton field generated by the string  $n'$ is taken on
the world-sheet of the string $n$: \be \lb{phi1}
\pa_\mu\stackrel{1}{\phi}|_{x=\stackrel{0}X_n(\tau,\si)}=\fr{\al\mu}{8\pi^2i}
\int
\fr{\e^{iq_\la(d^\la_{n'}-X^\la_{n}(\tau,\si))}\delta(qu_{n'})
\delta(q\Sigma_{n'})}{q^2+2i\ep q^0}\;q_\mu\;d^4 q.\ee
Substituting (\ref{phi1}) into (\ref{em11}}) one can obtain the
solution dividing the right hand side by the two-dimensional
D'Alembert operator  as follows: \ba\label{x1p}
\stackrel{\phi}{X}{\!\!}^{\mu}_{n }(\tau ,\sigma )=
i\fr{\al^2\mu}{(2\pi)^2} \int \frac{\Delta_{n'}[ q^\mu   +
\Sigma^\mu_n(q\Sigma_n) - u^ \mu_n(qu_n)] \e^{-iq_\la
\stackrel{0}{X}{\!\!}^\la_n(\tau,\si)}}{q^2[(qu_n)^2-(q
\Sigma_n)^2 ]} d^4 q , \ea where  \begin{equation}\label{delt}
\Delta_{n'}= \e^{iqd_{n'}}\delta(q u_{n'})\delta(q\Sigma_{n'}).
\end{equation} Note, that delta-functions in the integrand have
support outside both the light cone $q^2=0$ and the surface $(q
u_n)^2=(q\Sigma_n)^2$ (except for the trivial point $q^\mu=0$), so
the integral is finite.
\par
Consider now the two-form interaction (Fig. 3b). We have to solve
the equation \be\lb{em11a} (\pa^2_\tau
-\pa^2_\si)\stackrel{B}{X}{\!\!}^\mu_n =2\pi f V^{\nu\la}_n
\stackrel{1}{H}{\!\!}^{\mu}_{\;\nu\la}
|_{x=\stackrel{0}X_n(\tau,\si
 )}\, ,\ee where the  three-form is \be \lb{B1}
 \stackrel{1}{H}{\!\!}^\mu_{\,\, \nu\la}|_{x=\stackrel{0}X_n(\tau,\si)}
=\fr{f}{2\pi i} \int \fr{\Delta_{n'}q^{\{ \mu}V_{n'}^{\nu\la
\}}e^{-iq_\la \stackrel{0}X{\!\!}^\la _n(\tau,\si)}}{q^2+2i\ep
q^0}d^4 q,\ee where  curly brackets $\{\}$ denote the cyclic
permutation of indices. Again, dividing by the operator $
(\pa^2_\tau -\pa^2_\si) $ one obtains: \ba\label{x1B}
 \stackrel{B}{X}{\!\!}^{\mu}_{n }(\tau ,\sigma )= i\fr{f^2}{\mu} \int
\frac{{\Delta_{n'}V_{n \, \nu\la }q^{\{ \mu}V_{n'}^{\nu\la \}}}
\e^{-iq_\la
\stackrel{0}{X}{\!\!}^\la_n(\tau,\si)}}{q^2[(qu_n)^2-(q
\Sigma_n)^2]} d^4 q . \ea
\par
Similarly,  the gravitational  contribution is described by the
equation: \be\lb{em11h} (\pa^2_\tau
-\pa^2_\si)\stackrel{h}{X}{\!\!}^\mu_n =\fr{\mu}2 U^{\al\beta}_n
\left[\pa^\mu \stackrel{1}{h}_{\al\beta }-2\pa_\al
\stackrel{1}{h}{\!\!}^\mu_{\beta }-U^{\mu\nu}_n\left(\pa_\nu
\stackrel{1}h_{\al\beta}-2\pa_\beta \stackrel{1}h_{\al
\nu}\right)\right]_{x=\stackrel{0}X_n(\tau,\si
 )}\, ,\ee where the variation $\stackrel{1}h_{\mu\nu}$ is generated by the partner
string $n'$: \be \lb{h1}
\stackrel{1}{h}{\!\!}_{\mu\nu}|_{x=\stackrel{0}X_n(\tau,\si
 )}=\fr{4\mu G}{\pi} \int \fr{\Delta_{n'} W_{n'\mu\nu}
 e^{-iq_\la \stackrel{0}X{\!\!}^\la _n(\tau,\si)}}{q^2+2i\ep q^0}d^4
q.\ee Solving this equation one finds: \ba\label{x1h}
 \stackrel{h}{X}{\!\!}^{\mu}_{n }(\tau ,\sigma )= i\fr{2 \mu G}{\pi} \int
\frac{{\Delta_{n'}U_{n \,  \al\beta}}\left[ q^\mu
W^{\al\beta}_{n'}-2W^{\mu\al}_{n'} q^\beta -U^{\mu\nu}_n\left(
q_\nu W^{\al\beta}_{n'}-2q^\al W^{\beta }_{n'\nu }\right)\right]
\e^{-iq_\la
\stackrel{0}{X}{\!\!}^\la_n(\tau,\si)}}{q^2[(qu_1)^2-(q
\Sigma_1)^2]} d^4 q. \ea
\par It can be  checked that the gauge condition
(\ref{cond}) imposed in the beginning of the calculation holds
indeed for each of the three separate contributions to the
perturbed mapping functions.

\section{Effective sources of radiation }
The first order fields
$\stackrel{1}\phi,\;\stackrel{1}B{\!\!}^{\mu\nu},\;
\stackrel{1}h{\!\!}^{\mu\nu}$ do not contain the radiative parts.
Consider, e.g., the Fourrier-transform of the dilaton:\be
\lb{Furfi} \phi(k)=\int \phi(x)\e^{ikx}d^4x.\ee  The retarded and
advanced solutions of the first order wave equation (\ref{eeqop})
will read:
\begin{equation}\label{}
\stackrel{1}\phi{\!\!}^{\pm}_n(k)=  2\pi^2\mu\al
\fr{\delta(ku_n)\delta(k\Sigma_n)}{k^2\pm 2i\ep k^0}\e^{ikd_{n}}.
\end{equation}
The radiative part   \begin{equation}\label{rad}
\stackrel{1}\phi{\!\!}^{{\rm rad}}_n(k)=
\fr12\Big(\stackrel{1}\phi{\!\!}^{+}_n(k)-
    \stackrel{1}\phi{\!\!}^{-}_n(k)\Big)=-i\pi^3\mu\al
 {\delta(ku_n)\delta(k\Sigma_n)}\delta(k^2)\e^{ikd_{n}}.
\end{equation}
is the distribution having support only at the trivial point
$k^\mu=0$ in the momentum space. Thus, to investigate radiation,
we have to pass to the second order of the perturbation theory.
The problem reduces to the construction of the source terms in the
wave equations of the second order.
\subsection{Dilaton}
Consider the second order dilaton equation (\ref{sod}) in more
details. The current at the right hand side contains the
contributions localized on the string world-sheets (the upper line
in (\ref{sod})) and the bulk contributions coming from the
products of the first order dilaton, two-form  and  graviton
fields (the lower line). The former contains the sum over the
strings which can be understood as follows. One has to take the
perturbations of the world-sheet mapping functions $X^\mu_n$ for
each string $n=1, 2$ due to first order field generated by the
partner string $n=2, 1$ respectively. These contributions are
depicted by the graphs $a_n,\, b_n,\,c_n,\,n=1,2$ on the Fig. 4.
\begin{figure}
\centering
\includegraphics[width=6cm,height=6cm]{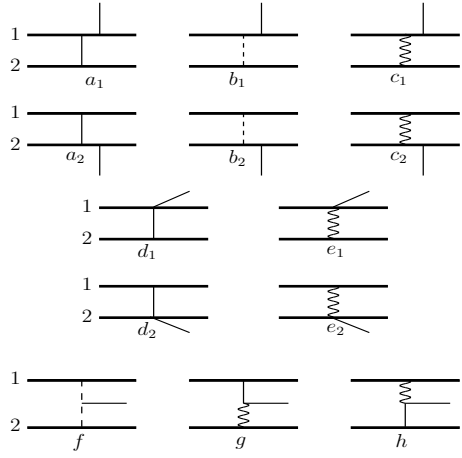}
\caption{The diagrams contributing to dilaton radiation in the
second order of the perturbation theory: $a_1,\, b_1, \,c_1$
correspond to deformation of the  first string and $a_2,\, b_2,
\,c_2$ -- to deformation of the second one; graphs $d,\,e $ stand
for contact terms. Diagram $f$ corresponds to the product of
(first order) two-form fields generated by the first and the
second string, graphs $g,\,h$ --- to mixed graviton-dilaton
contributions. }\label{f:4}
\end{figure}
The external dilaton leg corresponds to the emitted dilaton with
the momentum $k^\mu$ (in our classical treatment to the
Fourrier-transform of the current), thus the terms in the sum in
(\ref{sod})) with $n=1$ are given by the graphs $a_1, b_1, c_1$
and those with $n=2$ --- by the graphs $a_2, b_2, c_2$. Other
upper line terms must be treated in a similar way: one has to take
$U^{\mu\nu}_n$ for each $n=1, 2$ and multiply by the graviton and
dilaton perturbations caused by the other string $n=2, 1$. These
are depicted by the graphs $d_n,\; e_n,\; n=1,2$. In this way the
self-action of the fields upon the string will be excluded. \par
On the contrary, the terms in the lower line of the Eq.
(\ref{sod}) are non-local and not pairwise. Here the self-action
terms also have to be excluded, so we take in the quadratic
combinations only the products of the first order fields generated
by different strings.
\par
Consider first the contributions described by the graphs
$a_1,\,b_1,\,c_1$ (dilaton emission from the first string line).
The corresponding current reads: \be\lb{Jst1} \stackrel{st}J_1(x)
=- \fr{\mu\al}{8\pi}\int \stackrel{1}{X}{\!\!}_{1 }^\mu\pa_\mu
\delta^4\Big[x-\stackrel{0}{X}{\!\!}_{1}
(\tau,\sigma)\Big]\;d^2\si,\ee where the perturbation of the
mapping function of the first string $\stackrel{1}{X}{\!\!}^\mu_1$
is caused by the first order dilaton, two-form and graviton fields
generated by the second string. Substituting the corresponding
terms into (\ref{Jst1}) we obtain after some rearrangements:
\be\lb{x1mst} \stackrel{1}{X}{\!\!}^\mu_1 (\tau,\si) = i\int
\frac{Q^\mu_1\delta(qu_2)\delta(q\Sigma_2)
\e^{iq(d_2-d_1-u_1\tau-\Sigma_1\si)}}{q^2[(qu_1)^2-(q\Sigma_1)^2]}
d^4 q. \ee The vector $Q^\mu_1$ is the sum of three terms
according to the decomposition described above:  \be \lb{Q1}
Q_1^\mu=\fr{\al^2\mu}{(2\pi)^2}D_1^\mu +\fr{2 f^2}{\mu} Y^\mu_1
+\fr{2\mu G}{\pi}Z_1^\mu, \ee where the dilaton exchange
contribution is \be D_1^\mu = q^\mu+\Si^\mu_1 (q\Si_1) -u^\mu_1(q
u_1),\ee the two-form contribution is \ba Y^\mu_1&=&q^\mu
\left[(u_1u_{2})(\Sigma_1 \Sigma_{2})-( \Sigma_1 u_{2})(u_1
\Sigma_{2})\right]  +\Si_2^\mu \left[(qu_1)(u_{2}
\Sigma_1)-(\Sigma_1q)(u_1u_{2})\right] +\non\\&+& u_2^\mu
\left[(u_1 \Sigma_{2})( \Sigma_1q)-(qu_1)(\Sigma_1 \Sigma_{2})
\right],  \ea and the graviton contribution is \ba Z^\mu_1&=&
q^\mu \left[(u_1u_2)^2+(\Si_1\Si_2)^2-(u_1\Si_2)^2-(u_2\Si_1)^2-2
\right] +\left[u_1^\mu(qu_1)-\Si_1^\mu(q\Si_1)\right]\cdot\non\\
&&{\!\!}\left[(u_1u_2)^2-(\Si_1\Si_2)^2-(u_1\Si_2)^2+(u_2\Si_1)^2+
2\right] -2\left[u_1^\mu(q\Si_1)+\Si_1^\mu(q
u_1)\right]\left[(u_1u_2)(u_2\Si_1)-(\Si_1\Si_2)(u_1\Si_2)\right]-\non\\
&-&2u_2^\mu \left[(u_1u_2)(q u_1)-(\Si_1u_2)(q\Si_1)\right] +
2\Si_2^\mu\left[(u_1\Si_2)(q u_1)-(\Si_1\Si_2)(q\Si_1)\right].
 \ea
\par Consider the Fourier-transform \be \lb{Fur} J(k)=\int
J(x)\e^{ikx}d^4x \ee with $k^2=0$ (on-shell condition for the
massless particle). Substituting the above expressions and
integrating over the world-sheet of the first string one obtains
two more delta-functions: \be \int\e^{i(k-q)(u_1\tau-\Si_1\si)}
d\tau d\sigma =(2\pi)^2\delta[(k-q)u_1]\delta[(k-q)\Si_1],\ee so
totally we will have the product of four delta-functions in the
integrand: \be\lb{L1}
\Lambda_1(q,k)=\de(qu_1)\de(q\Si_1)\delta[(k-q)u_2]\delta[(k-q)\Si_2].\ee
Now consider the  contribution $J_2(k)$ coming from the second
string (the graphs $a_2, b_2, c_2$ on Fig.4). Obviously it can be
obtained from the previous result by interchanging indices $1
\leftrightarrow 2$ elsewhere. In this case we will get the product
of the $\de$ -functions  in the form \be\lb{L2}
\Lambda_2(q,k)=\de(qu_2)\de(q\Si_2)\delta[(k-q)u_1]\delta[(k-q)\Si_1].\ee
It is convenient to cast this second integral into the same form
as the previous one. For this it is sufficient to shift the
integration variable in $J_2(k)$ as follows: $q^\mu \rightarrow
k^\mu- q^\mu$. Since $\Lambda_1(q,k)=\Lambda_2(k-q,k)$, we will
get the same product of the $\de$-functions, so finally we can
present the total contribution of the first six graphs as follows
\be\lb{JST}\stackrel{st}J(k)=\stackrel{st}J_1(k)+\stackrel{st}J_2(k)=
\int\Pi(q,k)\left(\fr{\stackrel{st}\Theta_1(q)}{q^2}+
\fr{\stackrel{st}\Theta_2(k-q)}{(k-q)^2}\right)d^4q ,\ee where
\be\lb{PQK} \Pi(q,k)=\e^{i(d_1 k +q(d_2-d_1))}\Lambda_1(q,k), \ee
and \be\lb{ThST}\stackrel{st}\Theta_1(q)=\fr{\al}{[(q
\Sigma_1)^2-(qu_1)^2]}\left[\fr{\al^2 \mu^2}{8\pi}(k D_1)+\pi
f^2(k Y_1)+G\mu^2(k Z_1)\right].\ee To get the function
$\stackrel{st}\Theta_2$ from $\stackrel{st}\Theta_1$ one merely
has to interchange the indices $1,2$ labelling vectors in the
scalar products changing simultaneously $q^\mu\to (k-q)^\mu$.
\par
The contributions of the next four graphs $d_1,d_2,e_1,e_2$ are
computed similarly. The resulting  ``contact'' term in the source
can be presented again in the form (\ref{JST}) with equal
contributions from two strings
\be\lb{ThCT}\stackrel{ct}\Theta_1(q)=\stackrel{ct}\Theta_2(q)=
\fr{\al^3 \mu^2}{8\pi} +G\mu^2\al
\left[(u_1u_2)^2+(\Si_1\Si_2)^2-(u_1\Si_2)^2-(u_2\Si_1)^2-2
\right].\ee
\par Now consider the bulk terms (the second line in
(\ref{sod})) which are due to non-linear field interactions
dilaton-two-form and dilaton-graviton. Their contribution is
illustrated by the graphs $f$ and $e_n$, the second being
pairwise. In quantum theory terms they can be interpreted as the
coalescence of two virtual axions into the on-shell dilaton, and
the coalescence of the off-shell dilaton and graviton into the
on-shell dilaton. We have to compute the Fourrier-transform of the
bulk current \be\lb{JF}
\stackrel{b}J(x)=\fr{\al}{24\pi}\stackrel{1}{H}_{\mu\nu\la}
\stackrel{1}{H}{\!\!}^{\mu\nu\la}+
\fr1{4\pi}\partial_\mu\left[\left(\stackrel{1}{h}{\!\!}^{\mu\nu}-
\fr12\eta^{\mu\nu}\stackrel{1}{h}\right) \partial_\nu
\stackrel{1}{\phi} \right].\ee Here in the products
$\stackrel{1}{H}_{\mu\nu\la} \stackrel{1}{H}{\!\!}^{\mu\nu\la}$
and $\stackrel{1}{\phi} \stackrel{1}{h}_{\mu\nu}$ we have to
substitute one field generated by the first string and another by
the second one.
\par In spite of the different different appearance of the bulk terms
as compared to the world-sheet terms, one can cast them into the
unique form (\ref{JST}) as well. Consider the corresponding
transformations for  the first term in (\ref{JF}) depicted by the
graph $f$ on Fig. 4.  The first order quantities to be substituted
here read: \be \lb{H1}
\stackrel{1}{H}{\!\!}^{\mu\nu\la}_n(x)=\fr{f}{2i\pi} \int
\fr{\Delta_{n}q^{\{\mu} {V_{n}}^{\nu\la\}}\e^{-iqx}}{q^2}d^4 q.\ee
(We can omit the $\ep$-term in the denominator indicating the
position of the pole since the resulting integral does not depend
on its shift from the real axis). Now let us calculate the
contraction \ba \int \e^{ikx}\stackrel{1}{H}_{\mu\nu\la}
\stackrel{1}{H}{\!\!}^{\mu\nu\la}d^4x &=&- (2\pi f)^2\int
\fr{\Delta_{1}(q)\Delta_{2}(q')q_{\{ \mu} V_{1\nu\la \}}q'^{\{
\mu} {V_2}^{\nu\la \}}\e^{i(k-q-q')x}}{q^2q'^2}d^4q=\non\\
&=&\lb{FHH} - (2\pi f)^2\int \fr{\Delta_{1}(q)\Delta_{2}(k-q)q_{\{
\mu} V_{1\nu\la \}} (k-q)^{\{ \mu} {V_2}^{\nu\la \}}
\e^{i(k-q-q')x}}{q^2(k-q)^2}d^4 q. \ea We are interested in the
on-shell value of $k^\mu$, i.e. $k^2=0$. In this case one can
write  \be\lb{qkq} \fr1{q^2(k-q)^2}=\left( \fr1{(q-k)^2}-\fr1{q^2}
\right)\fr1{2kq}. \ee  Using this relation one can cast the above
contraction into the form (\ref{JST}) with  convention that the
argument $ q^\mu $ is used for the  $n=1 $ terms, and $(k-q)^\mu $
for the $n=2$ terms. \par Performing similar calculations for the
graphs $g_1, g_2$ and combining with the above, we  obtain the
total bulk term in the form (\ref{JST}) with $\Theta =
\stackrel{b}\Theta$:\ba \stackrel{b}\Theta_1=\pi f^2\al\left[
\fr{(Y_{1}k )}{(k q)}-(u_1u_{2})(\Sigma_1 \Sigma_{2})+( \Sigma_1
u_{2})(u_1 \Sigma_{2})\right] + \mu^2\al
G\fr{(ku_{2})^2-(k\Si_{2})^2+(ku_{1})^2-(k\Si_{1})^2}{(kq)},\ea
with the rule of getting $\stackrel{b}\Theta_2$ from
$\stackrel{b}\Theta_1$ the same as before. In obtaining this
expression we have used the delta-functions in the integrand
fixing $(qu_1)=(q\Si_1)=0$ and
$(qu_2)=(ku_2),\;(q\Si_1)=(k\Si_1).$

\subsection{Two-form}
The source term in the equation (\ref{fe}) for the second order
two-form field can be presented as the sum of ten graphs shown on
Fig.5. 
\begin{figure}
\centering
\includegraphics[width=6cm,height=4cm]{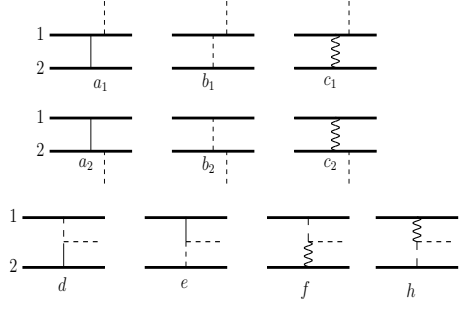}
\caption{The second order amplitudes of the two-form emission. The
graphs $a, b, c$ show contributions from  deformations of the
string world-sheets, the graphs $d, e, f, h$ --- the quadratic
bulk terms.}\label{f:5}
\end{figure}
Here we have string contributions corresponding to exchange
by the dilaton, two-form and graviton, the bulk terms describing
coalescence of virtual quanta to the on-shell axion, but no
contact term because of  absence of multi-leg vertices associated
with  two-form. The string terms $a_n, b_n, c_n$ are computed as
the Fourrier-transform of the currents \be\lb{JstB}
\stackrel{st}J_n^{\mu\nu}(x) =
f\int\left[\stackrel{0}{X}{\!\!}_{na}^\mu
\stackrel{1}{X}{\!\!}_{nb}^\nu\ep^{ab} -\fr12V_n^{\mu\nu}
\stackrel{1}{X}{\!\!}_{n}^\la\pa_\la\right]
\delta^4[x-X_n(\sigma,\tau)] d^2\si,\ee  where perturbations of
the mapping functions $X_n^\mu, n=1,2$ are generated by the
partner string $n=2,1$ respectively. Using the same rearrangements
as before, we obtain \be\lb{JSTB}\stackrel{st}J{\!\!}^{\mu\nu}(k)=
\stackrel{st}J{\!\!}_1^{\mu\nu}(k)+\stackrel{st}J{\!\!}^{\mu\nu}_2(k)=\int
\Pi(q,k)\left(\fr{\stackrel{st}\Theta{\!\!}_1^{\mu\nu}(q)}{q^2}+
\fr{\stackrel{st}\Theta{\!\!}_2^{\mu\nu}(k-q)}{(k-q)^2}\right)d^4q
,\ee where\ba\lb{ThSTB}\stackrel{st}\Theta{\!\!}_1^{\mu\nu}(q)=&-&
\fr{1}{[(q \Sigma_1)^2-(qu_1)^2]}\Bigg[8\pi^2
\fr{f^3}{\mu}\left((q \Sigma_1) u_1^{[\mu} {Y_{1}^{\nu ]}}- (q
u_1) \Sigma_1^{[\mu} {Y_{1}^{\nu ]}} -\fr12 {V}_1^{\mu \nu}
(Y_1k)\right)+\non\\&+&\fr12\al^2 f\mu\left((q \Sigma_1)
u_1^{[\mu} {D_1^{\nu ]}}- (q u_1) \Sigma_1^{[\mu} {D_1^{\nu ]}}
-\fr12 {V}_1^{\mu \nu} (D_{1} k) \right)+\\&+&8\pi f \mu G \left(
(q \Sigma_1) u_1^{[\mu} {Z_1^{\nu ]}}+ (q u_1) \Sigma_1^{[\mu}
{(Z_{1}^{\nu ]}} -\fr12 {V}_1^{\mu \nu} (Z_{1}k)
\right)\Bigg],\non \ea and to  get the second string term we must
interchange indices $1 \leftrightarrow 2$ and momenta $q^\mu
\leftrightarrow k^\mu - q^\mu$.
\par The bulk terms $d_n, e_n$ are the Fourrier-transform of the
bulk current \be\lb{JFB}
\stackrel{f}J(x)_{\mu\nu}=\fr{1}{4\pi}\partial_\la\left[
\stackrel{1}{H}{\!\!}^{\la}_{\,\,\,\,\mu\nu}
\left(4\al\stackrel{1}{\phi}{\!\!}-\fr12\stackrel{1}{h}{\!\!}
\right)\right],\ee where again we have to take the products of
fields generated by different strings. The result can be cast into
the form (\ref{JSTB}) with \ba\lb{TFB}
\stackrel{b}\Theta{\!\!}^{\mu\nu}_1&=&- \fr{8\pi f\mu G}{(k
q)}\Big[ {V}_1^{\mu \nu}(k q)+(q^{[\mu }{u_1}^{\nu ]}
)(\Si_{1}q)-(q^{[\mu } {\Si_1}^{\nu ]})(u_{1}q)  - {V}_{2}^{\mu
\nu}(k q)+(q^{[\mu }{u_2}^{\nu ]} )(\Si_{2}k)-(q^{[\mu }
{\Si_2}^{\nu ]})(u_{2}k)) \Big]-\non \\&-&  \fr{f \al^2\mu}{2(k
q)}\Big[{V}_1^{\mu \nu}( k q))-(q^{[\mu } {u_1}^{\nu ]}
)({\Si_1}q)+(q^{ [\mu }{\Si_1}^{\nu ]})(u_{1}q)- {V}_{2}^{\mu
\nu}(k q)-(q^{[\mu }{u_2}^{\nu ]})(\Si_{2}k)+(q^{[\mu
}){\Si_2}^{\nu ]}(u_{2}k) \Big], \ea and the same rule for the
second term.

\subsection{Graviton}
The source (\ref{taumn}) in the second order graviton equation can
be treated along the same line. It includes contributions of the
thirteen graphs shown on Fig. 6.
\begin{figure} \centering
\includegraphics[width=6cm,height=6cm]{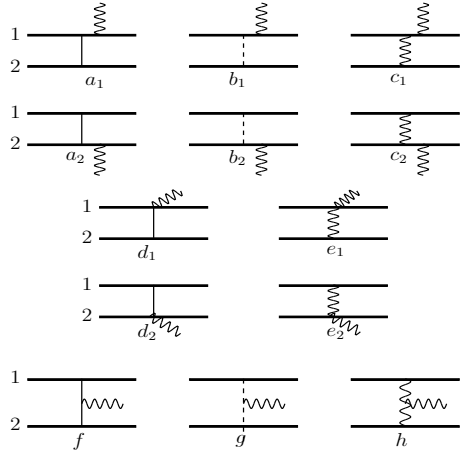}
\caption{The second order amplitudes of the graviton emission. The
sum of all graphs is zero on the graviton mass-shell.}\label{f:6}
\end{figure}It turns out, however, that the
projection of the total on-shell ($k^2=0$) current onto the
graviton transverse polarization states gives zero. The reason
lies in the dimensionality of the space transversal to the string:
the configuration of the parallel strings is reduced to that of
point particles in the 1+2 theory, where there is no on-shell
transverse gravitational degrees of freedom. As it was shown in
\cite{GaGrLe93}, the transformation to the parallel string
configuration is always possible for a superluminal moving
intersecting strings. So there is no gravitational radiation in
this case and we do not give here the details of the calculation.

\section{Radiation} Total radiation four-momentum loss can be
presented in a standard way through the on-shell Fourrier
transform of the source current, the vector $k^\mu=(\omega, {\bf
k}),\;\omega=|{\bf k}|$ playing the role of the radiation
four-momentum. In the case of the dilaton one obtains the
following explicitly Lorentz-covariant expression:\be
P^{\mu}_{(\phi)} \ = \ \fr{16}{\pi}\int k^\mu
\fr{k^0}{|k^0|}|J_{(\phi)}(k)|^2 \delta (k^2)d^4k, \ee and
similarly for the two-form \be\label{tot} P^{\mu}_{(B)} \ =  \
\fr1{\pi}\int k^\mu \fr{k^0}{|k^0|}|J_{\al\beta}(k)|^2 \delta
(k^2)d^4k.\ee Alternatively, the latter quantity can be presented
as a square of the polarization projection of the current. Indeed,
in three space dimensions the two-form  field propagating along
the wave vector $\bf k$ has a unique polarization state
\be\lb{pol} e_{ij}=\fr1{\sqrt2}(e_i^\theta e_j^\varphi-e_i^\varphi
e_j^\theta),\quad i, j=1,2,3,\ee where ${\bf e}^\theta$ and ${\bf
e}^\varphi$ are two unit vectors orthogonal to ${\bf k}$ and to
each other: \be {\bf e}^\varphi\cdot {\bf e}^\theta=0,\quad {\bf
k}\cdot {\bf e}^\varphi={\bf k}\cdot {\bf e}^\theta=0.\ee Using
antisymmetry and transversality of the two-form current $k^\mu
J_{\mu\nu}(k)=0$, and the completeness condition \be e_i^\theta
e_j^\theta + e_i^\varphi e_j^\varphi =\delta_{ij}-k_i
k_j/\omega^2,\ee one finds \be\label{totpol}  P^{\mu }_{(B)} =
\fr1\pi\int k^\mu \fr{k^0}{|k^0|}|J_{(B)}^{ij}(k)e_{ij}|^2 \delta
(k^2)d^4k.\ee \par Integrating over $k^0$, we finally obtain
\ba\lb{pp} P_{(\phi)}^{\mu } &=& \fr{16}{\pi}\int
\fr{k^\mu}{|{\bf k}|}|J_{(\phi)}({\bf k})|^2 d^3k,\\
 \label{totpol3}  P^{\mu }_{(B)} &=& \fr1\pi\int
\fr{k^\mu}{|{\bf k}|}|J_{(B)}({\bf k})|^2 d^3k, \ea where \be
\lb{jka}J_{(B)}({\bf k})=J_{(B)}^{ij}(k)e_{ij}\ee with $k^0=|{\bf
k}|$. In what follows we shall use the parametrization of
three-vectors by the spherical angles: $ {\bf k} =
\omega[\sin\theta \cos\varphi , \, \,  \sin\theta \sin\varphi, \,
\, \cos\theta],
 {\bf e}^\varphi =[-\sin\varphi, \, \, \cos\varphi, \, \, 0],
{\bf e}^\theta = [\cos\theta\cos\varphi, \, \,
\cos\theta\sin\varphi, \, \, -\sin\theta].$
\subsection{Cerenkov condition} With our conventions, the radiation
amplitudes associated with the first string contain the integral
\be\lb{Iq} I_1=\int\fr{\de (q u_2)\de (q \Sigma_2)\de [(k-q)
u_1]\de [(k-q) \Sigma_1]\,\,f(q)\e^{iqd}}{q^2}  d^4 q ,\ee where $
d^\mu=d_2^\mu - d_1^\mu $, and $ f(q) $ - is some regular function
of $ q $. In the Lorentz frame where $ u^\mu_1=(1,0,0,0),\,\,
\Sigma^\mu_1=(0,0,0,1),\,\, u^\mu_2=\ga \,
(1,0,-v\coal,v\sial),\,\, \Sigma^\mu_2=(0,0,\sial,\coal) $, one
can integrate over $ q^0 $ using \be\lb{d0} \de [(k-q) u_1]=\de
(q^0-\om). \ee Two other $ \de $-functions \ba\lb{d1} \de (q
u_2)&=&\ga^{-1} \de(\om +q_y v \cos\al -q_z v \sin\al), \\ \lb{d2}
\de (q\Sigma_2)&=&\ga^{-1} \de(q_y  \sin\al + q_z v \cos\al),\ea
can be used to fix the values of $q_y, q_z$: \be\lb{ad2}
q_y=-\fr{\om\cos\al}{v}\, ,\quad\quad q_z=\fr{\om\sin\al}{v}\,.\ee
The remaining $ \de $-function \be\lb{rd} \de [(k-q) \Sigma_1]=\de
(k_z-q_z)=\de\left(k_z-\fr{\om \sin\al}{v} \right)\ee no more
depends on $q$, but rather restricts the value of the wave vector
of radiation $k_z$: \be\lb{k z} k_z=\fr{\om \sin\al }{v} =v_p \ee
This is  Cerenkov's condition for an emitted massless wave.
Indeed, in our frame an effective source of radiation is moving
along z-axis with the velocity $ v_p $, thus the quantity
\be\lb{cot} \cos\theta =\fr{k_z}{\om}=\fr1{v_p} \ee defines
Cerenkov's angle of emission if $ v_p > 1 $, i.e. the source is
super-luminal. Thus, radiation arises if the string relative
velocity $ v $ and the inclination angle $ \al $ satisfy the
Cerenkov condition \be\lb{Cc} \sin\al < v .\ee Given $v$, this
condition will always be satisfied for sufficiently small $\al$.
In particular, it is identically fulfilled for $\al =0$, i.e. for
parallel strings. Moreover, it can be shown that, if $v_p >1$, the
Lorentz frame and the world sheet coordinates  $\tau_n,\sigma_n$
always exist such that  two string look parallel \cite{GaGrLe93}.
In the frame where the first string is at rest and streched along
the $z$-axis, an effective source moves in $z$-direction, and
radiation will be emitted along the Cerenkov cone around the
$z$-axis with an angle \be \theta_{\rm Ch}= \arccos
\frac{v}{\sial}.\ee\par The remaining integral over $ q_x $ in
(\ref{Iq}) can be evaluated using contour integration. With our
choice of coordinates, the scalar product $q^\mu d_\mu =-q_x d $,
so we have the integral over $q_x$ \be\lb{Iqx} I_1=\int
\fr{\e^{-iq_x d} f(q_x) dq_x}{q_x^2+p^2}
=\fr{\pi}{p}f(-ip)\,\e^{-\ka d},\ee where\be
p=\sqrt{q_y^2+q_z^2-q_0^2}=\fr{\om}{\ga v}.\ee For the second term
in (\ref{JST})  containing the pole $(k-q)^{-2}$  one obtains the
same fixed values of $q^0, q_y, q_z$, but the integration over
$q_x$ gives the value  \be q_x=k_x+\fr{\om \xi}{i v},\quad
\xi=\cos\al+v\sin\theta\sin\phi.\ee \par Summarizing the above
results, we obtain: \ba \lb{iden}
\int\left(\frac{\Theta_1(q)}{q^2}+\frac{\Theta_2(k-q)}{(k-q)^2}\right)
\e^{ikd_1 +iqd_2} \; d^4q=
\left[E_1\Theta_1(q_1)+(\ga\xi)^{-2}E_2\Theta_2(k-q_2)
\right]\de(\cos\theta-\cos\theta_{\rm Ch}), \ea where \be\lb{EE}
E_1=\e^{ikd_1-\om d/(v\ga)},\quad E_2=\e^{ikd_2-\om d\xi/v},\ee
and the following complex vectors are introduced: \ba\label{qnew}
{q_1}^\mu &=& \fr{\om}{v} \, [ v,\, -i/\ga,\,-\cos\al,\,\sin\al ], \\
k^\mu-{{ q_2^\mu }}&=&\fr{\om\xi}{v} \, [ 0,\,i,\,1,\,0]. \ea
\par The presence of the delta-function at the right hand side of
(\ref{iden}) means that the total radiation loss is infinite. This
could be expected since we deal with the stationary motion of an
infinitely long strings. So it is natural to consider the
radiation loss per unit length of the string at rest. Redefining
the currents as \be J({\bf k})=I({\bf
k})\de(\cos\theta-\cos\theta_{\rm Ch}),\ee and using the identity
\be \de^2(\cos\theta-\cos\theta_{\rm Ch})=
\frac{L\om}{2\pi}\de(\cos\theta-\cos\theta_{\rm Ch}),\ee where $L$
is the normalization length, we find \ba\lb{dilperlen} {\cal
P}_{(\phi)}^{\mu} &=&L^{-1} P_{(\phi)}^{\mu}=\fr{8 }{\pi^2}\int
 {k^\mu}|I_{(\phi)}({\bf k})|^2 \de(\cos\theta-\cos\theta_{\rm Ch})d^3k,\\
 \label{axperlen}  {\cal P}^{\mu}_{(B)}&=& L^{-1} P^{\mu}_{(B)}
 =\fr{1}{2\pi^2}\int
 k^\mu  |I_{(B)}({\bf k})|^2\de(\cos\theta-\cos\theta_{\rm Ch})
d^3k, \ea
\subsection{Relativistic peaking and spectrum enhancement}
According to (\ref{iden}), there is a frequency cut-off due to
exponential factors (\ref{EE}). Actually the amplitude is
superposition of two terms which can be associated with
contributions of two strings (this can not be taken literally: in
the second order of the perturbation theory the superposition
principle does not hold). Two terms exhibit different frequency
cut-off. The first sting term has a cut-off \be\lb{om1}\om\lesssim
\fr{v\ga}{d},\ee which does not depend on the radiation angle,
while the second one exhibits  a $\varphi-$dependent cut-off:
\be\lb{om2}\om\lesssim
\fr{v}{d\xi}=\fr{v}{d(\cos\al+v\sin\theta\sin\varphi)}.\ee This
means that the angular distribution of radiation on the Cerenkov
cone is anisotropic.  This feature becomes especially pronounced
in the ultrarelativistic case.
\par
In view of the identity \be\cos^2\al=v^2\sin^2\theta+\ga^{-2},\ee
which holds on the radiation cone,  in the ultrarelativistic limit
$\ga\to\infty$ the quantity $\xi$ has a sharp minimum at
$\varphi=-\pi/2$ corresponding to the direction of the moving
string in the rest frame of the first string:
\be\lb{xiultra}\xi\thickapprox
\fr{1}{2\ga^2\cos\al}\left(1+\beta^2\ga^2\cos\al^2\right)],
 \ee where
$\beta=\pi/2+\varphi\ll 1$. Due to the factor $\xi^{-2}$ in the
second term in (\ref{jka}) the Cerenkov radiation is peaked around
the direction $\varphi=-\pi/2$ within the narrow angular region
\be\lb{bega} \beta\lesssim \ga^{-1}.\ee Moreover, the frequency
range associated with the second terms is substantially larger in
the ultrarelativistic limit than that associated with the first
term. Indeed, if  $\ka=\ga\cos\al\gg1,$ one has: \be
\xi\approx\frac{1}{2\ga\ka}\left(1+\ka^2\beta^2\right),\ee so the
frequency range extends up to the
frequency\be\om\lesssim\fr{\ga\ka}{d}\ee in the angular region
(\ref{bega}). Therefore, radiation exhibits relativistic peaking
in the forward direction in the same way as radiation of the
ultrarelativistic particle. This could be expected, since
relativistic peaking has purely kinematical nature.

\subsection{Dilaton radiation}
Collecting the world-sheet contributions to radiation amplitudes
we obtain after integration over $q$:
\ba\lb{stdil}\stackrel{st}I(k)+\stackrel{ct}I(k)=\fr {\al}{\om^2}
\Bigg\{\fr{\al^2\mu^2}{8}\left({E_1}
\left[1-\fr{i\cos\phi+\ka\sin\phi}{\ga v\sin\theta}\right] +\fr{
E_2}{\ga^2\xi^2} \left[\fr{i\e^{-i\varphi}}{\ga v \sin\theta} +
 2\xi\ga \right]\right)+ \non\\ + \pi^2 f^2\left( E_1
\fr{\sin\phi+i \ka\cos\varphi} {\ga v\sin\theta}
-\fr{E_2}{\ga^2\xi^2} \fr{\ga v\sin\theta+i
\ka\,\e^{-i\varphi}}{\ga v \sin\theta}\right)+
\\+ \mu^2\, G\,\pi\left( E_1 [( \ka\sin\varphi \!-
\!i\cos\phi)+\ga v \sin\theta ]\! +\!\fr{ E_2}{\ga^2\xi^2}
ie^{i\varphi} \right)\ga v \sin\theta\Bigg\}.  \ea Here the upper
line corresponds to  the dilaton exchange, the second line
--- to the two-form and the last line to the graviton exchange.
The bulk contribution after integration reads:
 \ba\stackrel{b}I(k)= \fr{\pi^2 f^2\al}{\ga \xi} \left[
E_1  (\ga \ka\xi -1-i\ga v \sin\theta\cos\varphi)  +  iE_2 \ga
v\sin\theta\e^{i\phi} \right] +\non \\ + \fr{G \mu^2 \pi}{\ga \xi}
\left[ E_1 (\ga v \sin\theta + \ka\sin\phi-i\cos\phi) + E_2i
\e^{i\phi} \right]2\ga \ka v \sin\theta.\ea Consider first the
Cerenkov's threshold $v=\sial$, when the radiation cone shrinks to
$\theta_{\rm Ch}=0$. This corresponds to $\ka=1,\;\xi=1/\ga$.
Since in this limit $kd=0$, the exponents become equal $E_1=E_2$
and the bulk term vanishes. The graviton contribution to
(\ref{stdil}) vanishes too, while the dilaton and two-form
contributions differs only by the coefficients. The total
radiation amplitude at the threshold will be given by
\be\lb{CJF1A} I_{\rm thr}(k)=\fr{\al}{8\om^2}E_1 \left(3
\al^2\mu^2 -8\pi^2 f^2\right). \ee Integrating the expression
(\ref{dilperlen}) for $\mu=0$ (the energy loss rate) over the
angles in $d^3k=\om^2 d\om d\cos\theta d\varphi$, we obtain the
infrared-divergent integral over frequencies. Introducing the
inverse correlation distance $\Delta$  as an infrared cutoff
parameter one finds:\be {\cal P}_{\rm thr}
=\frac{\al^2}{4\pi}(3\al^2\mu^2-8\pi^2f^2)^2
\int_{\Delta^{-1}}^\infty \fr{d\om}{\om}\exp({-2\frac{\om
d}{v\ga}}).\ee  In the BPS limit (\ref{BPS}) one has \be {\cal
P}_{\rm thr} =\frac{\mu^4\al^6}{ \pi} \int_{\Delta^{-1}}^\infty
\fr{d\om}{\om}\exp({-2\frac{\om d}{v\ga}}).\ee  Integrating this
over frequencies we obtain the total radiation rate  in terms of
the integral exponential function (see Appendix): \be {\cal
P}_{\rm thr}= \frac{\mu^4\al^6}{\pi} {\rm
Ei}\left(1,\frac{2d}{v\ga\Delta}\right).\ee For small impact
parameters  $d\ll v\ga\Delta$ this expression can be approximated
by the logarithm  \be  {\cal P } \approx \frac{\mu^4\al^6}{\pi}\ln
\left(\frac{v\ga\Delta}{2d\e^C}\right),\ee where $C$ is the Euler
constant, $\e^C=1.781072418$. For large impact parameters, $d\gg
v\ga\Delta$, radiation exponentially falls off:\be {\rm Ei}
\left(1, \frac{2d}{v\ga\Delta}\right)\approx \frac{2d}{v\ga\Delta}
\exp\left(-\frac{2d}{v\ga\Delta}\right). \ee
\par
For $\theta\neq 0$ the radiation amplitudes are more complicated.
Note, that in our reference frame the first string is at rest
while the second one is moving. For this reason the radiation
amplitudes are not symmetric with respect to two strings. As it
was observed in the previous section, in the ultrarelativistic
case the frequency range associated with the factor  $E_2$ is much
larger than that that with $E_1$, and in this limit the $E_2$
terms are dominant. Collecting the dominant terms we obtain for
the total amplitude of the dilaton emission in the
ultrarelativistic case \ba\label{jDc} I({\bf k})= \fr{\al
E_2}{\ga^2\xi^2\om^2} \Big\{ \fr{\al^2\mu^2}8
\left(\fr{i\e^{-i\phi}}{\ga v\sin\theta} +2\ga\xi\right)- {\pi^2
f^2 } \left( 1 + \fr{i \cos\al \,\e^{-i\phi}}{v\sin\theta}- i\xi
\ga^2 v \sin\theta \, \e^{i\phi} \right)+\non\\+
    {\mu^2 G\pi}i \ga v\sin\theta (1+2 \ka\ga\xi)\e^{i\phi}
\Big\}.\ea Substituting this into (\ref{dilperlen}) and taking
into account the angular peaking near $\phi=-\pi/2$  we obtain the
spectral-angular distribution of radiation per unit length in the
vicinity of this direction: \be \lb{Pkapbom1}\fr{d{\cal P} }{d\om
d\beta}= \fr{32\al^2\kappa^2}{\om}
\fr{(\Omega_1+\Omega_2\ka^2\beta^2)^2}{(1+\ka^2\beta^2)^4}
\exp\left({-\fr{\om d(1+\kappa^2\beta^2)}{\ga\kappa}}\right),\ee
where  \be\lb{k12}
  \Omega_1=4 G\mu^2\ka^2+\pi f^2\ka\quad
  \Omega_2= 2G\mu^2\ka^2+ \pi f^2\ka+ \fr{\al^2\mu^2}{4\pi}.\ee
 Dividing by $\omega$ we will get also the number of emitted
 dilatons \begin{equation}\label{numd}
    \fr{d{\cal N}}{d\om d\beta}=\frac1{\om}\fr{d{\cal P} }{d\om
d\beta}.
\end{equation}
Note, that the graviton, axion and dilaton exchange terms entering
into the above expressions for the parameters $\Omega_{1,2}$
exhibit different behavior in the invariant Lorentz factor $\ka$,
the dominant as $\ka\to\infty$ being the gravitational term.
\par The spectrum exhibits an infrared divergence, and in the
forward direction $\beta=0$ it extends up
 to $\om\sim\om_{{\rm max}}$, where \be  \om_{{\rm
 max}}=\fr{\ga\kappa}{d}.\ee Integrating over frequencies
with the infrared cut-off $\Delta^{-1}$   we obtain the angular
distribution of  radiation: \be\lb{Pkapa}  \fr{d{\cal P} }{
d\beta}= 32\al^2 \kappa^2
 \fr{(\Omega_1+\Omega_2\ka^2\beta^2)^2}{(1+\ka^2\beta^2)^4}{\rm  Ei}
\left(1, \fr{d(1+\beta^2\ga\kappa)}{\ga\kappa\Delta}\right) .\ee
Since the integral exponential function decays exponentially for
large argument, the total radiation is peaked around $\beta=0$
within the angle \be\beta\lesssim \sqrt{\ga\kappa}.\ee One can
also obtain the spectral distribution of radiation extending the
integration domain over $\beta$ in (\ref{Pkapbom1}) to the full
axis in view of the exponential decay of the integrand: \be
\lb{Pom1} \fr{d{\cal P}}{d\omega}= \om\fr{d{\cal
N}}{d\om}\fr{2\pi\al^2 d}{3\ga}
[\Omega_1^2F_1(z)+2\Omega_1\Omega_2F_2(z)+ \Omega_2^2F_3(z)],\ee
where $z=\fr{\om d}{\ga \ka}$ and the functions $F_i$ are
expressed in terms of the probability integral
(\ref{er}):\ba\lb{Fs} F_1(z)&=&(8z^2-16z+30)\fr{\e^{-z}}{\sqrt{\pi
z}}-(8z^2-12z+18-\fr{15}{z}) {\rm erfc}(\sqrt{z}) ,\non\\
F_2(z)&=&(-8z^2-8z+6)\fr{\e^{-z}}{\sqrt{\pi
z}}+(8z^2+12z-6+\fr{3}{z}) {\rm erfc}(\sqrt{z})  ,\\
\lb{F3} F_3(z)&=&(8z^2+32z+6)\fr{\e^{-z}}{\sqrt{\pi
z}}-(8z^2+36z+18-\fr{3}{z})  {\rm erfc}(\sqrt{z})  .
 \non\ea
For small  frequencies $\omega\ll \ga\kappa/d$ these functions
grows as \be F_1(z)\sim\fr{15}{z},\quad F_2(z)\sim \fr3{z},\quad
F_3(z)\sim \fr3{z}, \ee while for large $z$ they exponentially
decay: \be F_1(z)\sim \fr{48\e^{-z}} {\sqrt{\pi z^3}},\quad F_2(z)
\sim -\fr{57\e^{-z}}{2\sqrt{\pi z^5}},\quad F_3(z) \sim
\fr{105\e^{-z}}{2\sqrt{\pi z^5}}.\ee To regularize the infrared
divergence one has to introduce the cut-off length $\Delta$. After
integration over frequencies $ \om $ from $\Delta^{-1}$ to
infinity we obtain the total dilaton radiation rate in the
ultrarelativistic limit:\be \lb{TotD1} {\cal P} =\fr{ 2\pi\al^2
\kappa}{3}\left[(\Omega_1^2f_1(y)+2\Omega_1\Omega_2f_2(y)+\Omega_2^2f_3(y))
\right],\ee where three new functions are introduced \ba
f_1(y)&=&5f(y)+{\rm
erfc}(\sqrt{y}))\left(\fr83y^3-6y^2+18y+\fr{37}{2}\right)-
\fr{\e^{-y}\sqrt{y}}{\pi}\left( \fr83y^2-\fr{22}{3}y+23 \right),
\\ \lb{f2def}f_2(y)&=&f(y)-{\rm
erfc}(\sqrt{y})\left(\fr83y^3+6y^2-6y-\fr{5}{2}\right)+
\fr{\e^{-y}\sqrt{y}}{\pi}\left( \fr83y^2+\fr{14}{3}y-7 \right), \\
f_3(y)&=&f(y)+ {\rm
erfc}(\sqrt{y})\left(\fr83y^3+18y^2+18y+\fr{1}{2}\right)-\fr{\e^{-y}\sqrt{y}}{\pi}\left(
\fr83y^2+\fr{50}{3}y+ 11 \right), \ea where \be
y=\frac{d}{\ga\ka\Delta},\ee and the function $f(y)$ is expressed
through the generalized hypergeometric function (\ref{hyp})\be
f(y)=12\sqrt{\fr{y}{\pi}}\,
{_2F_2}\left(\fr12,\fr12;\fr32,\fr32;-y\right)
-3\ln\left(4y\e^{C}\right).\ee For small $y$  these functions grow
as \be f_1(y)\sim 15\ln{\fr1y}, \quad f_{2,3}(y)\sim 3\ln{\fr1y},
\ee while for large $y$ they tend to zero in view of the
asymptotic relation (\ref{ashyp}). \par Similar expressions can be
obtained for the total number of dilatons:\be \lb{TotND1} {\cal N}
=\fr{ 2\pi\al^2
d}{3\ga}(\Omega_1^2\cF_1(y)+2\Omega_1\Omega_2\cF_2(y)+
\Omega_2^2\cF_3(y)), \ee where three more functions are introduced
\ba \cF_1(y)&=& {\rm erfc}(\sqrt{y} )\left(
4y^2-12y-39+\fr{15}{y}\right)-\fr{\e^{-y}}{\pi\sqrt{y}}\left(
4y^2-14y-30 \right)-6f(y),\\ \cF_2(y)&=& {\rm erfc}(\sqrt{y})
\left(-4y^2-12y-9+\fr{3}{y}\right)+\fr{\e^{-y}}{\pi\sqrt{y}}\left(
4y^2+10y+6 \right)-2f(y), \\ \cF_3(y)&=&  {\rm erfc}(\sqrt{y})
\left(4y^2+36y+9+\fr{3}{y}\right)-\fr{\e^{-y}}{\pi\sqrt{y}}\left(
4y^2+34y-6 \right)-6f(y), \ea tending to zero at infinity, and \be
\cF_1\sim\fr{15}{y},\quad \cF_{2,3}\sim\fr{3}{y} \ee for small
$y$.
\par
As we have noted, in the ultrarelativistic limit the dominant
contribution to radiation amplitude comes from the graviton
exchange term. Leaving only this contribution we find for the
radiation rate:
 \be \lb{Totp1} {\cal P}^{(\phi)}=\fr{8}{3}\pi
 G^2\al^2\mu^4\ka^5 g(y) , \ee where
 \be \lb{gydef} g(y)=25f(y)+ {\rm
erfc}(\sqrt{y})\left(\fr83y^3-30y^2+114y+\fr{169}{2}\right)-
\fr{\e^{-y}\sqrt{y}}{\pi}\left( \fr{8}3y^2-\fr{94}{3}y+131
\right). \ee
\begin{figure}
\centering
\includegraphics[width=6cm,height=6cm,angle=-90]{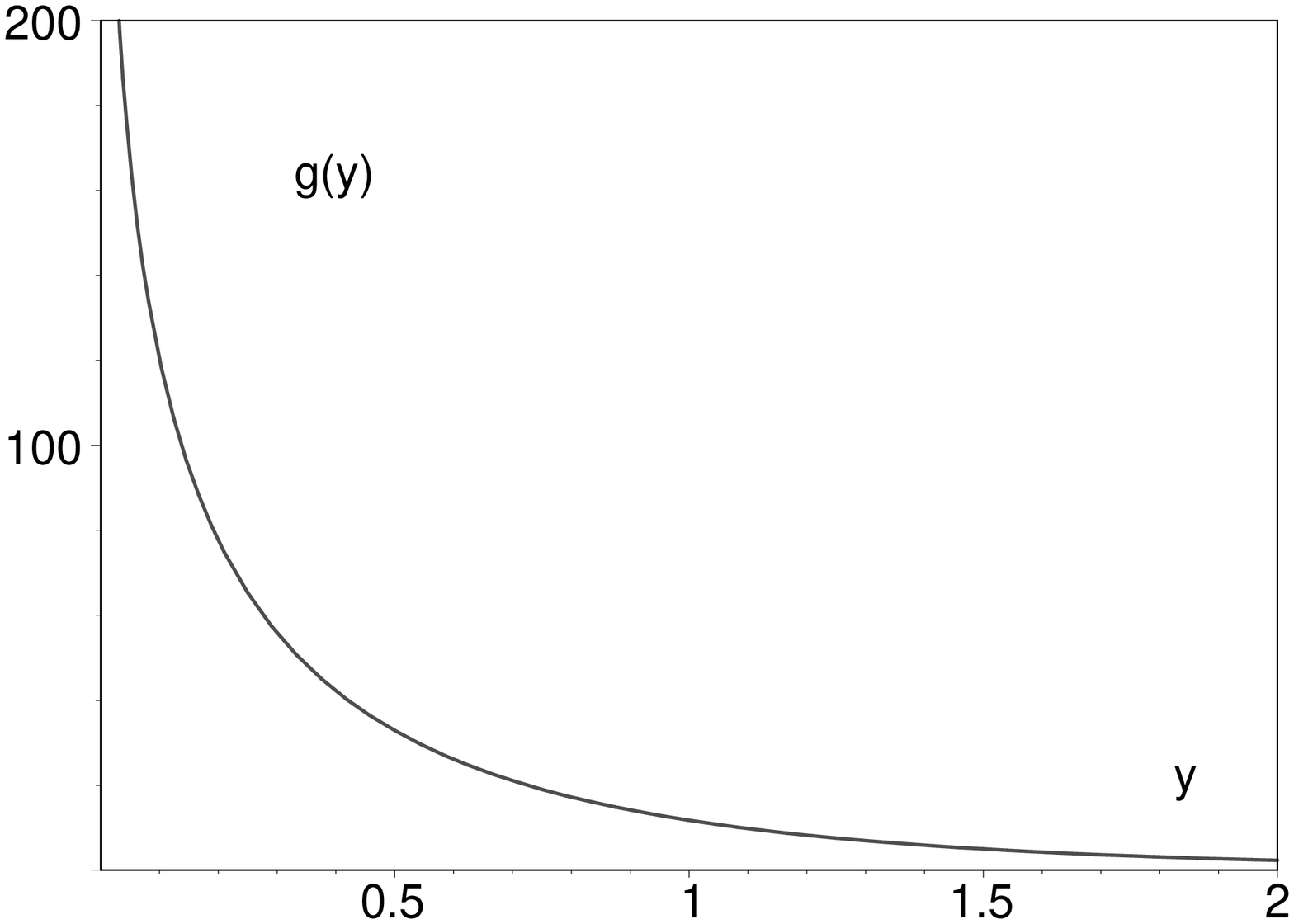}\hspace{12mm}
\centering
{\includegraphics[width=6cm,height=6cm,angle=-90]{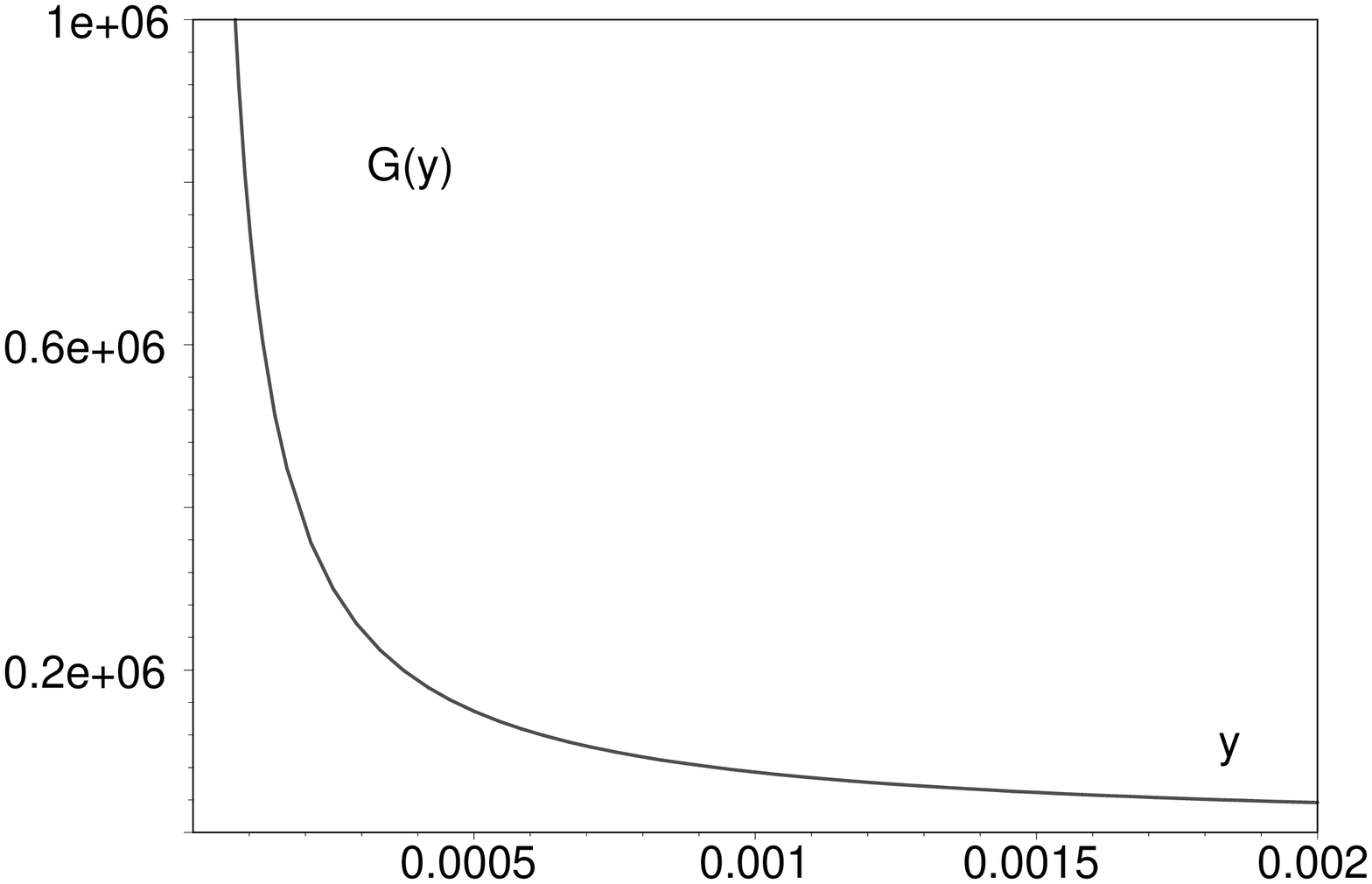}}\hspace{12mm}
\\
\parbox[t] {0.47\textwidth}
{\caption{g(y).}\lb{f:7}}
\parbox[t] {0.47\textwidth}{\caption{$G(y)$.}\lb{f:8}}
\end{figure}
Similary, for the dilaton number: \ba \lb{TotN1} {\cal
N}^{(\phi)}&=&\fr{8d}{3\ga}\pi
 G^2\al^2\mu^4\ka^4{\cal G}(y),\\
  {\cal G}_(y)&=&{\rm
erfc}(\sqrt{y}))\left(+4y^2-60y-183+\fr{75}{y}\right)-
\fr{\e^{-y}}{\pi\sqrt{y}}\left( 4y^2-62y-150 \right)-38f(y). \ea
The corresponding numerical curves are shown on the
Figs.\ref{f:7},\ref{f:8}. For small $y$ one has: \ba {\cal
P}^{(\phi)}&=&200\pi
 G^3{\bar\al}^2\mu^4\ka^5 \ln\left(\fr{\ga\ka\Delta}{d}\right) ,\\
 {\cal N}^{(\phi)}&=&200\pi
 G^3{\bar\al}^2\mu^4 \ka^5 \Delta,\ea
where we introduced the dimensionless dilaton coupling constant.
These quantities rapidly grow with increasing Lorentz factor of
the collision $\ga$ (recall that $\ka=\ga\cos\al$ where $\al$ is
the angle between the strings). Thus, Cerenkov radiation is
greatly enhanced for ultrarelativistic velocities.
\subsection{Two-form radiation}
After integration over the momentum $q$, the string contribution
to the two-form radiation amplitude (\ref{JSTB}) takes the form:
\ba\stackrel{st}I(k)= \fr{ 4\sqrt{2}\pi^3 f^3}{\mu\ga
v\sin\theta}\left(  E_1 (\cos\varphi\!-\!i \ka\sin\varphi) + \fr{
E_2}{ \ga^2\xi^2}
\left[(i v\xi\ga^2\sin\theta -\e^{-i\varphi}\right)\right)+ \non  \\
+\fr{\al^2 f\mu \pi\sqrt{2}}{2{\ga v\sin\theta}}\left(  E_1
(i\sin\varphi- \ka\cos\varphi)+\fr{ E_2}{\ga^2\xi^2 }\left(
\ka\e^{-i\varphi }-i v\ga\sin\theta -2\ga^3\xi v^2
\sin^2\theta\cos\varphi \right)\right)+ \non  \\+4\sqrt{2}\pi^2 f
\mu G \ga v\sin\theta\left( E_1 ( \ka\cos\varphi +
i\sin\varphi)-\fr{ E_2}{\ga^2\xi^2} \left[i\ga v \sin\theta +
\ka\e^{i\varphi}-2\cos\varphi\ga\xi\right] \right). \ea The bulk
amplitude (\ref{JSTB}) will read: \ba\stackrel{b}I(k)=
\left(\fr{\al^2 f\mu\pi \sqrt{2}}{2}+
 4\sqrt{2}\pi^2 f \mu G \ga v\sin\theta\right)
\left(\fr{ E_1}{\ga \xi} [ (1-2\ga\xi)\ga
v\sin\theta\cos\varphi-i] \!+\!\fr{ E_2}{ \xi }(2
v\sin\theta\cos\varphi -i \xi )
 \right). \ea
Consider first the Cerenkov threshold $v=\sial$. In the above
expressions the graviton exchange term in the string term
vanishes, and the bulk term is zero. The remaining string
amplitudes due to the dilaton and the two-form exchange  simplify
as follows: \be\lb{JB2A}I_{\rm thr}(k)=\fr{i\pi f
}{\sqrt{2}\mu\om^2}E_1 \, \left(8\pi^2f^2 - 3\al^2\mu^2\right).
\ee Substituting this into (\ref{axperlen}) and integrating over
the  angles we obtain the spectral distribution of the two-form
radiation on the Cerenkov  threshold: \be  {\cal P}_{\rm thr}
=\fr{\pi f^2}{2\mu^2}\left(8\pi^2f^2 - 3\al^2\mu^2\right)^2
\int_{\Delta^{-1}}^\infty \fr{d\om}{\om} \exp\left({-2\frac{\om
d}{v\ga}}\right),\ee where $\Delta$ the infrared cutoff frequency.
After integration over frequencies one finds: \be {\cal P}_{\rm
thr} = \frac{4(2\pi)^5 f^6}{\mu^2} {\rm
Ei}\left(1,\frac{2d}{v\ga\Delta}\right).\ee For small impact
parameters  $d\ll v\ga\Delta$ this expression can be approximated
by
 \be    {\cal P}_{\rm thr}\approx\frac{4(2\pi)^5 f^6}{\mu^2}\ln
\left(\frac{v\ga\Delta}{2d\e^C}\right),\ee for $d\gg v\ga\Delta$
one can use \be {\rm Ei} \left(1,
\frac{2d}{v\ga\Delta}\right)\approx \frac{2d}{v\ga\Delta}
\exp\left(-\frac{2d}{v\ga\Delta}\right). \ee
\par Now consider the case of the arbitrary Cerenkov angles
in the ultrarelativistic limit. Leaving only the relativistic
string contribution, we obtain: \ba\label{jfc} I({\bf k})=
 \fr{fE_2}{\ga^2\xi^2\om^2}
\Big\{\fr{4\sqrt{2}\pi^3 f^2}{\mu \ga v\sin\theta}(i\ga^2 \xi
v\sin\theta-i \e^{-i\phi}) +\fr{\pi \mu\al^2}
 {\sqrt{2}}\left(\fr{ \ka\,\e^{-i\phi}+\ga
v\sin\theta}{\ga v \sin\theta}-i\ga^2\xi^2 \right)- \non\\ -
 {4\sqrt{2}\pi^2 G\mu} [  i(1- \ka^2) + \ka\ga
v\sin\theta e^{i\phi})-i\ga^2\xi^2]\Big\}. \ea Substituting this
into (\ref{axperlen}) we obtain the spectral-angular distribution
in the vicinity of $\phi=\-\pi/2$   \be\lb{PkapbomB} \fr{d{\cal P}
}{d\om d\beta}=\om\fr{d{\cal N} }{d\om d\beta}=\fr{64\pi^4
\kappa^2f^2}{ \om}
\fr{\chi_B^2(1-\kappa^2\beta^2)^2+4\kappa^4\beta^2(\chi_h\ka^2
-\chi_D)^2}{(1+\kappa^2\beta^2)^4} \exp\left({-\fr{\om
d(1+\kappa^2\beta^2)}{\ga\kappa}}\right),\ee where \be
\chi_D=\frac{\mu \al^2}{8\pi^2},\quad \chi_B=\frac{f^2}{\mu},\quad
\chi_h=\frac{G\mu  }{\pi}.\ee Integrating over the angles we get:
\ba \lb{Pom2} \fr{d{\cal P} }{d\omega} &=&\fr{16\pi^5 d}{3\ga}
[F_4(z)\chi_B^2+F_2(z)(\chi_h\ka^2 -\chi_D)^2],  \\ \lb{Nom2}
\fr{d{\cal N} }{d\omega} &=&\fr{16\pi^5 d^2 f^2}{3\ga^2\ka z}
[F_4(z)\chi_B^2+F_2(z)(\chi_h\ka^2 -\chi_D)^2],\ea where \be
F_4=(8z^2+8z+6)\fr{\e^{-z}}{\sqrt{\pi z}}-(8z^2+12z+6-3/z) {\rm
erfc}(\sqrt{z}),  \ee  and $F_2$ is given by (\ref{Fs}). For small
and large frequencies  one has: \be F_4(z)\sim\fr{3}{z},\quad
F_4(z)\sim \fr{12\e^{-z}} {\sqrt{\pi z^3}}\ee respectively.
Finally, integrating over frequencies we find the total two-form
radiation rate and the number of axions: \ba \lb{TotB1} {\cal
P}^{(B)}&=&\fr{16\pi^5 \kappa f^2}{3}
\left(\chi_B^2f_4(y)+(\chi_h\ka^2 -\chi_D)^2f_2(y)\right),\\
\lb{TotNB1} {\cal N}^{(B)}&=&\fr{16\pi^5 d
f^2}{3\ga}\left(\chi_B^2\cF_4(y)+(\chi_h\ka^2
-\chi_D)^2\cF_2(y)\right),\ea where \ba f_4(y)&=&f(y) +{\rm
erfc}(\sqrt{y})\left(\fr83y^3+6y^2+5y+\fr{7}{2}\right)-\fr{\e^{-y}\sqrt{y}}{\pi}\left(
\fr83y^2+\fr{14}{3}y+ 5 \right),\\ \cF_4(y)&=& {\rm erfc}(\sqrt{y}
)\left(4y^2+12y-3+\fr{3}{y}\right)-\fr{\e^{-y}}{\pi\sqrt{y}}\left(
4y^2+10y-6 \right)-2f(y). \ea For small $y$   \be
\cF_4\sim\fr{3}{y},\quad f_4(y)\sim 3\ln{\fr1y}, \ee  while for
large $y$  both functions tend to zero. \par The result of the
calculation of the axion Cerenkov radiation in the flat space-time
\cite{GMK04} is reproduced putting $\chi_D=\chi_h=0.$ It reads \be
{\cal P}_0^{(B)}=\frac{16\pi^5 \ka f^6}{3\mu^2} f_4(y).\ee   In
our case the dominant contribution comes form the graviton
exchange. In the BPS limit one has
\be\chi_D=\chi_B,\quad\chi_h=\frac{8}{{\bar\al}^2}\chi_B,\ee where
$\bar\al$ is the dimensionless dilaton coupling constant.  For
large $\ka$ the leading term is \be\lb{TotB2}{\cal
P}^{(B)}=\frac{2^{10}\pi^5 \ka^5 f^6}{3{\bar\al}^4\mu^2}
f_2(y).\ee In the most interesting case of small $y\; f_2\simeq
f_4$, so the the ratio \be \frac{{\cal P}^{(B)}}{{\cal
P}_0^{(B)}}=\frac{64 \ka^4}{{\bar\al}^4}\ee can be large, e.g. for
$\bar\al =1$ and $\ka=5$ it is equal to $4\cdot10^{7}$. \par  The
dominant term for the number of axions is \be \lb{TotNB2} {\cal N}
=\frac{2^{10}\pi^5 \ka^4 f^6 d}{3\ga{\bar\al}^4\mu^2} {\cal
F}_2(y).\ee The numerical curves $f_2(y),\,\cF_2(y)$ are shown on
Figs. \ref{f:9}, \ref{f:10}.
\begin{figure}
\includegraphics[width=6cm,height=6cm,angle=-90]{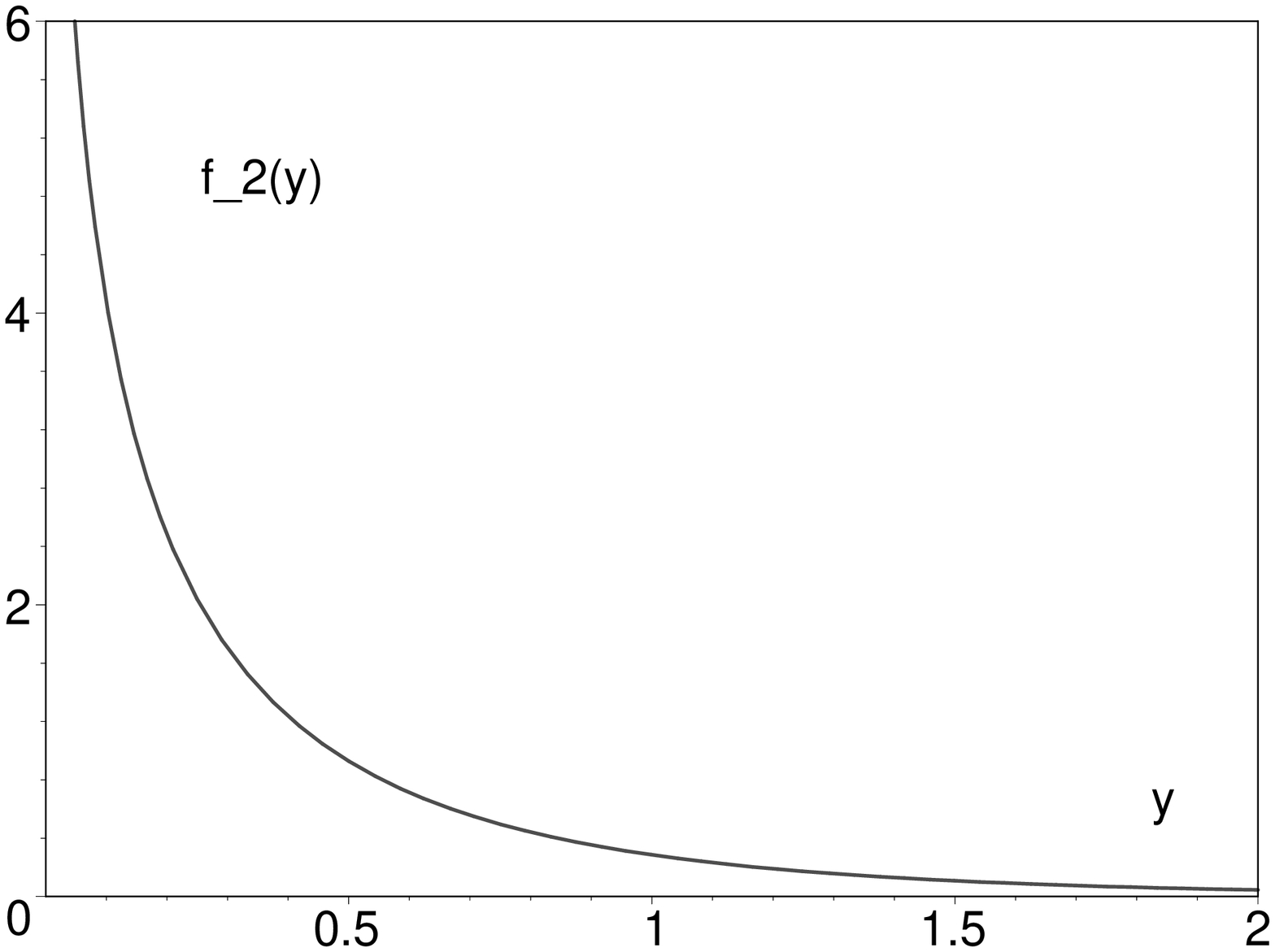}\hspace{12mm}
{\includegraphics[width=6cm,height=6cm,angle=-90]{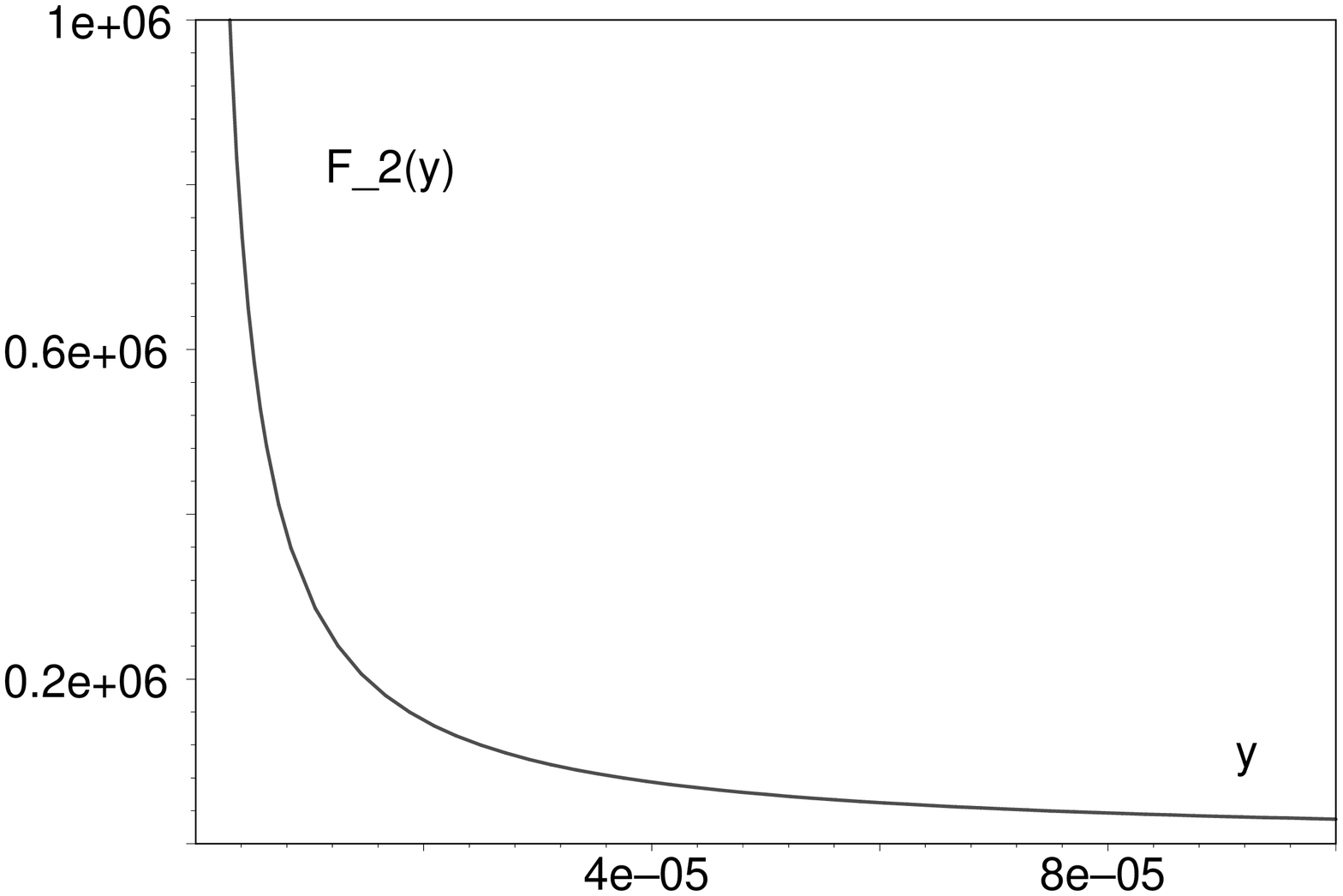}}\hspace{12mm}
\\
\parbox[t] {0.47\textwidth}
{\caption{$f_2(y)$.}\lb{f:9}}
\parbox[t] {0.47\textwidth}{\caption{${\cal F}_2(y)$.}\lb{f:10}}
\end{figure}
For small $y$ one has:\ba\lb{TotB2s}{\cal
P}^{(B)}&=&\frac{2^{10}\pi^5 \ka^5 f^6}{{\bar\al}^4\mu^2}
\ln\left(\frac{\ga\ka\Delta}{d}\right),\\ \lb{TotNB2s} {\cal N}
&=&\frac{2^{10}\pi^5 \ka^5 f^6 \Delta}{{\bar\al}^4\mu^2}.\ea

\section{Cosmological estimates}
Cosmic strings are formed as a network of long strings of the size
comparable to the horizon size. Colliding strings intercommute and
form closed loop. At some stage the scale-invariant string network
is formed consisting of long strings and loops which move freely
with relativistic velocities. Evolution of cosmic superstring
networks was recently discussed in
 \cite{Sa04}-\cite{Ty05}. In many respects cosmic
superstrings  are similar to the gauge theory cosmic strings, with
some distinctions, however.  In particular, for gauge theory
strings the probability of the formation  of loops $P$ is  of the
order of unity, whereas for F-strings $P \sim 10^{-3}$ and for and
D-strings $P \sim 10^{-1}$. The cosmic superstring network has a
scaling solution and the characteristic scale  is proportional to
the square root of the reconnection probability. A typical
separation between two long strings is comparable to the horizon
size $t$ (we use the standard cosmological units), $ \zeta (t)
\simeq \sqrt{P} t $. The results of numerical simulation show that
the network of long strings   reaches an energy density \be\lb{ed}
\rho_s = \fr\mu{\sqrt{P} t^2}. \ee
\par Let us estimate the energy loss of long strings due to
Cerenkov radiation of dilatons and axions. Consider an ensemble of
randomly oriented straight strings moving chaotically in space.
Let choose one target string between them and introduce the
Lorentz frame where it is at rest. Other strings will have
different orientations and velocities, and we can characterize
them very roughly as moving in three orthogonal directions with
equal probability. Since the dependence of the Cerenkov radiation
on the inclination angle $\al$ is smooth, we can use for an
estimate the particular result obtained for parallel strings
($\al=0$) introducing an effective fraction $\nu$ of ``almost''
parallel strings taking into account the effect of the angular
spread. Assuming $N$ to be the number of strings in the
normalization volume $V=L^3$, we have to integrate the radiation
energy released $\cal P$ in the collision with the impact
parameter $d=x$ over the plane perpendicular to the target string
with the measure $N/L^2\cdot 2\pi x dx$. To estimate the radiation
power per unit time we then have to divide the integrand by the
impact parameter. Multiplying this quantity by the total number of
strings $N$ to get the radiation energy released per unit time
within the normalization volume, we obtain for the Cerenkov
luminosity: \be Q_{C}=\int_0^L {\cal
P}\,\nu\,\fr{N}{L^2}\,\fr{N}{V}\,2\pi dx.\ee
For BPS strings we use as ${\cal P}$ the leading relativistic
terms (\ref{Totp1}) and (\ref{TotB2}). Taking into account that
the string number density is related to the energy density
(\ref{ed}) via \be \fr{N}{V}=\fr{\rho_s}{\mu L},\ee and assuming
for a rough estimate $L\sim\Delta\sim t$, where $t$ is
cosmological time, we obtain \ba Q^{(\phi)}_{C}
&\simeq &\fr{16}{3}\pi^2 G^3{\bar\al}^2\mu^4\ga^7 \nu S_1(w) \fr1{{P} t^3},\\
Q^{(B)}_{C} &\simeq&  \fr{2^{11}\pi^6 \ga^7 f^6  \nu}{3
{\bar\al}^4 \mu^2}S_2(w)\fr1{{P} t^3},\lb{QQ}\ea where \be
S_1(w)=\int_0^w g(y) dy,\quad S_2(w)=\int_0^w f_2(y) dy, \quad
w=\fr{L}{\ga^2\Delta}. \ee The exact values of these integrals are
given in the Appendix and shown on Figs.\ref{f:11},\ref{f:12}.
\begin{figure}
\includegraphics[width=6cm,height=6cm,angle=-90]{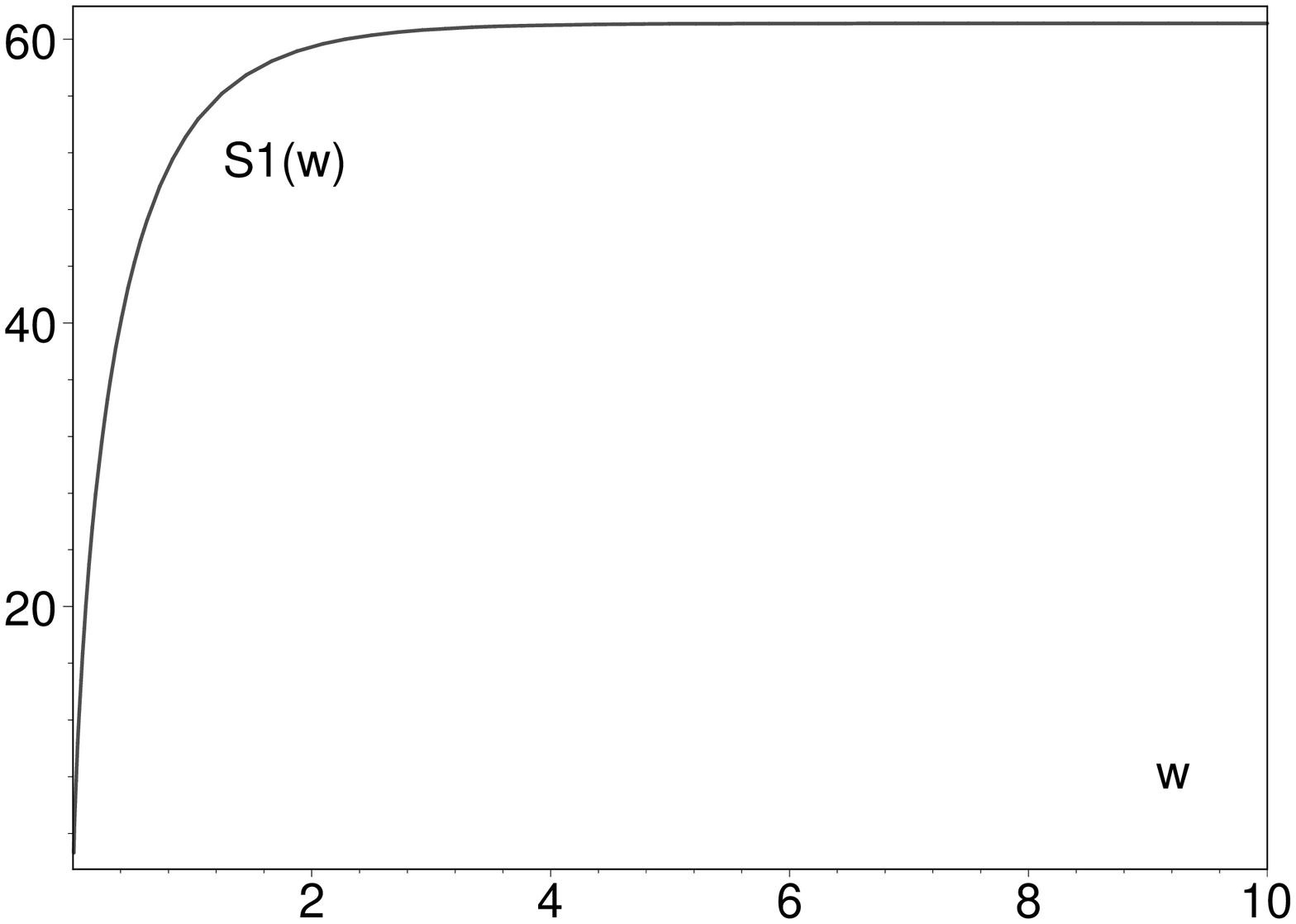}\hspace{12mm}
{\includegraphics[width=6cm,height=6cm,angle=-90]{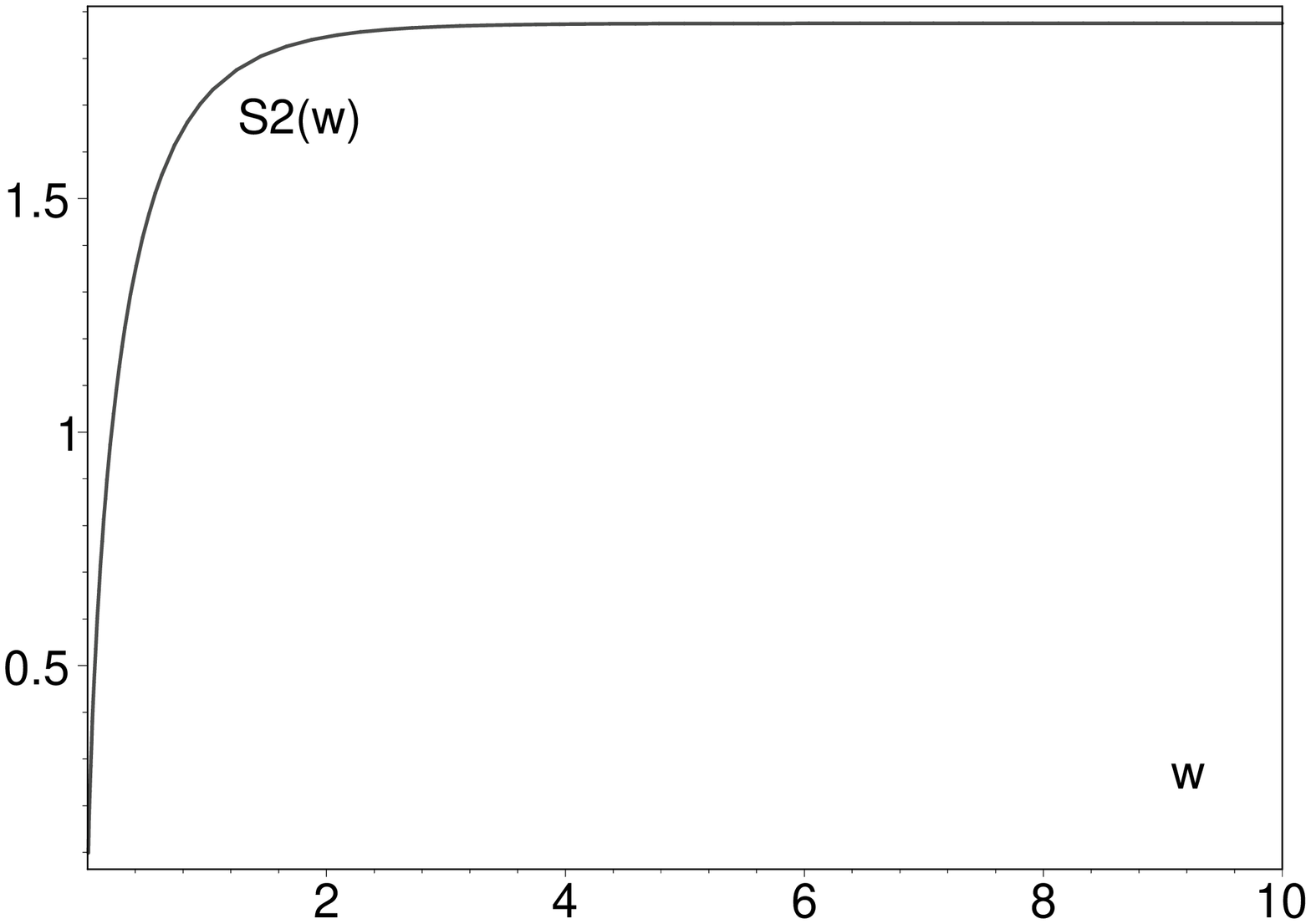}}\hspace{12mm}
\\
\parbox[t] {0.47\textwidth}
{\caption{ $S_1(w),\,w=L/(\ga^2\Delta)$}\lb{f:11}}
\parbox[t] {0.47\textwidth}{\caption{  $S_2(w) $}\lb{f:12}}
\end{figure}
Note that the
realistic value of $\ga$ is of the order of unity, so applying our
formulas obtained in the limit $\ga\gg 1$ is only an order of
magnitude estimate.
\par Now we can calculate the energy density $\va_C$ of Cerenkov
radiation as a function of time in the radiation dominated
Universe. The energy density of massless fields  scales with the
Hubble parameter $H$ as $H^{-4}$, so we have to solve the equation
\be \fr{d \va_C}{dt}=-4H\va_C+Q_{C},\ee where $H=\frac1{2t}$. Thus
we obtain for the energy density of the Cerenkov dilaton and
two-form radiation:
\ba  \va_C^{(\phi)} &\simeq& \fr{16}{3}\pi^2 G^3{\bar\al}^2\mu^4\ga^7 \nu S_1(w) \fr{\ln{(t/t_0)}}{{P} t^2} ,\\
\va_C^{(B)} &\simeq& \fr{2^{11}\pi^6 \ga^7 f^6  \nu}{3 {\bar\al}^4
\mu^2}S_2(w)\fr{\ln{(t/t_0)}}{{P} t^2} ,\ea  where $t_0$ - the
initial time of the long string formation.  Finally, using the
approximate formulas (\ref{S1sm},\ref{S2sm}) valid in the
relativistic case, we find:\ba  \va_C^{(\phi)} &\simeq& 800\pi^2
G^3{\bar\al}^2\mu^4\ga^5\ln\ga \,\nu \;\fr{\ln{(t/t_0)}}{{P} t^2} ,\\
\va_C^{(B)} &\simeq& \fr{2^{12}\pi^6 f^6\ga^5\ln\ga   \nu}{
{\bar\al}^4 \mu^2}\; \fr{\ln{(t/t_0)}}{{P} t^2} .\ea Comparing the
last expression for axions with the flat space result \cite{GMK04}
we observe an enhancement due to higher power of the Lorentz
factor and bigger numerical coefficient. This difference is due to
the fact that here the main contribution to the first order
interaction between the strings comes from gravitational force
which is proportional to energy. In view of the previous analysis
\cite{GMK04} we can conclude that Cerenkov radiation from long
strings is non-negligible effect in the cosmic superstring
network. More detailed analysis will be given elsewhere.

\section{Conclusion}
In this paper we have studied in detail Cerenkov radiation from
moving straight strings interacting with dilaton, two-form and
gravity. Formation of the faster-than-light sources in the system
of randomly oriented moving straight strings is rather generic. As
we have shown, these sources have a collective nature and arise
due to deformations of the strings world-sheets caused by their
interactions via massless fields. These deformations propagate
with superluminal velocities if the inclination angle is
sufficiently small, for parallel strings the source velocity is
infinite. Radiation wave vectors lie on the Cerenkov cone in the
same way as in the case of Cerenkov radiation of point charges in
dielectric media.

One interesting feature related to  dimensionality of a string
compared to a point charge is the absence of gravitational
radiation in four space-time dimensions. This is related to the
fact that the space transverse to the straight string is
two-dimensional, so the emitted massless fields must live in 1+2
dimensional space-time rather than in four-dimensional. As it is
well-known, gravity in 1+2 dimensions does not contain free
gravitons, this is why one can expect gravitational radiation from
straight strings to vanish. In higher dimensions this objection
does not work, so Cerenkov gravitational radiation can be expected
in space-time dimensions higher than four. In four dimensions
Cerenkov mechanism works for the dilaton and the two-form field
which is equivalent to a pseudoscalar.

To avoid complications due to possibility of ``physical'' string
intersections (leading to the well studied processes of
intercommutation and formation of loops) we consider the
``collision'' of strings moving in parallel planes. At each
instant of time there exists a point of minimal separation between
the strings, and it is this point which may propagate with the
superluminal velocity. When interaction between the strings via
dilaton, two-form and gravity is taken into account, strings get
deformed in the vicinity of this point, these deformations
contribute to an effective radiation source. Another contribution
comes from tensions associated with the first order fields which
give rise to these deformations. The string deformations give
contributions localized on the world-sheets, while the field
stresses give bulk contributions. Both have the same order of
magnitude.

Cerenkov radiation from strings has some peculiar features in the
highly relativistic region. We have shown that in this case
radiation  exhibits strong beaming on the Cherenkov cone in the
direction of the fast string in the rest frame of the target
string. The main radiation frequency is proportional to the
inverse impact parameter, but in the ultrarelativistic case the
spectrum is enhanced  to  high frequencies proportional to the
square of the Lorentz-factor of the collision. It is shown that in
this limit gravitational interaction between string dominates  and
gives the main contribution to the effective sources of dilaton
and two-form radiation.

Cerenkov's mechanism can be regarded as an analog of the
bremsstrahlung of point charges in electrodynamics, which gives
the main contribution to radiation in plasma. In the string case,
however, there is another radiation mechanism due to existence of
the internal string dynamics: radiation from oscillating loops.
This effect is of the first order in couplings between the string
and massless fields. Cerenkov radiation arises only in the second
order in these coupling, so presumably it is less important. But
as we have shown here, it has  stronger dependence on the Lorentz
factor of the string collision, so it must become dominant for
highly relativistic strings. Also, in the cosmic string network it
is a pairwise effect which gives contribution to the radiation
loss  proportional to the square of the density of strings. Our
rough cosmological estimates indicate that Cerenkov radiation is a
non-negligible effect in the cosmic string context indeed.

\section*{Acknowledgments}
This work  was  supported by RFBR grant 02-04-16949.

\section{Appendix}
Here we collect properties of some special functions used in the
main text.
\subsection{Integral exponential function}
The definition:\begin{equation}\label{ei}
 {\rm Ei}(1,z)=\int\limits_z^\infty\frac{\e^t dt}{t}.
\end{equation}
The series expansion:
\begin{equation}\label{eis}
{\rm Ei}(1,z)= -C-\ln z+z-z^2/2+z^3/18-\ldots ,
\end{equation}
where $C$ is Euler constant, $\e^C=1.781072418$. \par The
asymptotic expansion:
\begin{equation}\label{eia}
  {\rm Ei}(1,z)=  \frac{\e^{-z}}{z}\left(1-\frac1{z}+\frac2{z^2}+\ldots.\right) .
\end{equation}
\subsection{Probability integral}
 The definition: \begin{equation}\label{er}
{\rm erf}(z)=  \frac{2}{\sqrt{\pi}}\int\limits_0^z\e^{-t^2} dt.
\end{equation} The series expansion:
\begin{equation}\label{ers}
{\rm erf}(z) =
\frac{2z}{\sqrt{\pi}}\left(1-z^2/3+z^4/10-\ldots\right).
\end{equation}
The asymptotic expansion:
\begin{equation}\label{era}
  {\rm erfc}(z)\equiv 1-{\rm erf}(z)=  \frac{\e^{-z^2}}{\sqrt{\pi}z}
  \left(1-\frac1{2z^2}+\frac{3}{4 z^2}-\ldots\right).
\end{equation}
In Sec. 7 we  used the following indefinite integral:
\begin{equation}\label{indint}
    \int {\rm erf}(\sqrt{z})\frac{dz}{z}=4\sqrt{\frac{z}{\pi}}
{_2F_2}\left(\fr12,\fr12;\fr32,\fr32;-z\right),
\end{equation}
where
\be\lb{hyp}{_2F_2}\left(\al_1,\al_2;\beta_1,\beta_2;x\right)=
\sum_{k=0}^{\infty}\frac{\al_{1k},\al_{2k}}{\beta_{1k},\beta_{2k}}
\frac{x^k}{k!},\quad \al_k=\al(\al+1)\cdots(\al+k),\ldots\ee  is
the generalized hypergeometric function. To obtain an asymptotic
behavior of the latter for $z\to \infty$ we use the following
identity:\be \int\limits_0^\z \ln{w} \fr{\exp{(-w)}}{\sqrt{\pi w}}
dw =\ln z \;{\rm erf}(\sqrt{z}) - 4\sqrt{\frac{z}{\pi}}
{_2F_2}\left(\fr12,\fr12;\fr32,\fr32;-z\right),\ee which can be
easily proved integrating by parts. Then taking into account the
integral \be\int_0^\infty \ln{w} \fr{\exp{(-w)}}{\sqrt{\pi w}} dw
= -C-2\ln{2},\ee we find that for $z\to\infty$ \be\lb{ashyp}
4\sqrt{\fr{z}{\pi}}\,
{_2F_2}\left(\fr12,\fr12;\fr32,\fr32;-z\right)
\approx\ln\left(4z\e^{C}\right).\ee Integration of the functions
$g(y)$ defined in (\ref{gydef}) and $f_2(y)$ defined in
(\ref{f2def}) over $y$ can be performed analytically: \ba
S_1(w)&=&\int_0^w g(y) dy=300\sqrt{\fr{w}{\pi}} \Biggl[ \
_2F_2\left(-\fr12,- \fr12; \fr12, \fr32;-w\right)- {_2F_2}\left(-
\fr12,- \fr12; \fr12, \fr12;-w\right)\Biggr]\non\\&+&75w \left(
1-\ln\left(4w\e^{C}\right)\right)
+(\fr{169}{2}w+\fr23w^4+57w^2-10w^3-\fr{189}{8}){\rm
erfc}(\sqrt{w})-\non\\&-&\fr{1}{12}\fr{\e^{-y}\sqrt{y}}{\sqrt{\pi}}\left(8w^3-
124w^2+750w+ 567 \right)+\fr{189}{8}, \lb{gw} \\S_2(w)&=&\int_0^w
f_2(y) dy=12\sqrt{\fr{w}{\pi}} \Biggl[ \ _2F_2\left(-\fr12,-
\fr12; \fr12, \fr32;-w\right) - {_2F_2}\left(- \fr12,- \fr12;
\fr12, \fr12;-w\right)\Biggr]\non \\&+& 3w
\left(1-\ln\left(4w\e^{C}\right)\right)
+\fr1{24}(60w-16w^4+72w^2-48w^3-9) {\rm erfc}(\sqrt{w})+\non
\\&+&\fr{1}{24}\fr{\e^{-y}\sqrt{y}}{\sqrt{\pi}}\left(16w^3+
40w^2-84w-18 \right)+\fr38. \lb{f2w}\ea For small arguments the
leading
terms are \ba \lb{S1sm} S_1(w)&\simeq& -75w\ln w,\\
\lb{S2sm}S_2(w)&\simeq& -3w\ln w.\ea

\end{document}